\documentclass{aastex631}
\usepackage{CJK}
\usepackage{amsmath}

\newcommand\Jz{$J_{z}$}
\newcommand\Jzsp{$J_{z}$ }
\newcommand{\teff}{\ensuremath{T_{\textrm{eff}}}}
\newcommand{\teffsp}{\ensuremath{T_{\textrm{eff} } } }

\newcommand{\lnjz}{\ensuremath{\ln J_z}}
\newcommand{\lnjzsp}{\ensuremath{\ln J_z} }

\newcommand{\tgrid}{\ensuremath{\tau_{\textrm{grid}}}}
\newcommand{\tgridsp}{\ensuremath{\tau_{\textrm{grid}} } }

\newcommand{\lnjzgrid}{\ensuremath{\ln J_{z \textrm{grid}}}}
\newcommand{\lnjzgridsp}{\ensuremath{\ln J_{z \textrm{grid} } } }

\begin{document}

\title{\texttt{zoomies}: A tool to infer stellar age from vertical action in \textit{Gaia} data}

\author[0000-0002-3022-6858]{Sheila Sagear}
\affiliation{Center for Computational Astrophysics, Flatiron Institute, 162 5th Avenue, Manhattan, NY, USA}
\affiliation{Department of Astronomy, University of Florida, 211 Bryant Space Science Center, Gainesville, FL 32611}

\author[0000-0003-0872-7098]{Adrian M. Price-Whelan}
\affiliation{Center for Computational Astrophysics, Flatiron Institute, 162 5th Avenue, Manhattan, NY, USA}

\author[0000-0002-3247-5081]{Sarah Ballard}
\affiliation{Department of Astronomy, University of Florida, 211 Bryant Space Science Center, Gainesville, FL 32611}

\author[0000-0003-4769-3273]{Yuxi (Lucy) Lu}
\affiliation{American Museum of Natural History, Central Park West, Manhattan, NY, USA}

\author[0000-0003-4540-5661]{Ruth Angus}
\affiliation{American Museum of Natural History, Central Park West, Manhattan, NY, USA}
\affiliation{Center for Computational Astrophysics, Flatiron Institute, 162 5th Avenue, Manhattan, NY, USA}

\author[0000-0003-2866-9403]{David W. Hogg}
\affiliation{Center for Cosmology and Particle Physics, Department of Physics, New York University, 726 Broadway, New York, NY 10003, USA}
\affiliation{Max-Planck-Institut für Astronomie, Königstuhl 17, D-69117 Heidelberg, Germany}
\affiliation{Center for Computational Astrophysics, Flatiron Institute, 162 5th Avenue, Manhattan, NY, USA}

\begin{abstract}

Stellar age measurements are fundamental to understanding a wide range of astronomical processes, including galactic dynamics, stellar evolution, and planetary system formation. However, extracting age information from main-sequence stars is complicated, with techniques often relying on age proxies in the absence of direct measurements. The \textit{Gaia} data releases have enabled detailed studies of the dynamical properties of stars within the Milky Way, offering new opportunities to understand the relationship between stellar age and dynamics. In this study, we leverage high-precision astrometric data from \textit{Gaia} DR3 to construct a stellar age prediction model based only on stellar dynamical properties; namely, the vertical action. We calibrate two distinct, hierarchical stellar age--vertical action relations, first employing asteroseismic ages for red giant branch stars, then isochrone ages for main-sequence turn-off stars. We describe a framework called \texttt{zoomies} based on this calibration, by which we can infer ages for any star given its vertical action. This tool is open-source and intended for community use. We compare dynamical age estimates from \texttt{zoomies} with age measurements from open clusters and asteroseismology. We use \texttt{zoomies} to generate and compare dynamical age estimates for stars from the Kepler, K2, and TESS exoplanet transit surveys. While dynamical age relations are associated with large uncertainty, they are generally mass-independent and depend on homogeneously measured astrometric data. These age predictions are uniquely useful for large-scale demographic investigations, especially in disentangling the relationship between planet occurrence, metallicity, and age for low-mass stars.

\end{abstract}


\section{Introduction} \label{sec:intro}

The \textit{Gaia} data releases \citep{prusti_gaia_2016, vallenari_gaia_2023} have revolutionized our understanding of Galactic dynamics, along with stellar and planetary astronomy within a broader Galactic context. The motions of stars within the galaxy encode age information: generally speaking, older stars exhibit higher velocity dispersion or vertical action as a population \citep{wielen_diffusion_1977, rocha-pinto_chemical_2004, almeida-fernandes_method_2018}. Historically speaking, low-mass dwarfs have been especially useful in calibrating stellar age and Galactic kinematics, by virtue of their long lifetimes \citep{reid_palomarmsu_1995, faherty_brown_2009, kiman_exploring_2019, angus_exploring_2020, popinchalk_evaluating_2021}. 

Understanding the ages of stars is key to understanding not only how stars themselves evolve, but also how galaxies evolve and how planetary systems behave over time. Stellar ages are particularly interesting for constraining the occurrence rate and evolution of planetary systems over time. Transit surveys enabled by NASA's \textit{Kepler} Mission \citep{borucki_kepler_2010} and the Transiting Exoplanet Survey Satellite (\textit{TESS}, \citealt{ricker_transiting_2014}) have discovered thousands of exoplanets in the past decade \citep{guerrero_TESS_2021}. The number of exoplanets known today has not only enabled the study of individual planets and systems, but also their ensemble properties. Here \textit{Gaia} has also played a pivotal role by improving constraints on stellar properties (e.g. \citealt{fulton_california-kepler_2018, berger_gaia-kepler_2020, berger_gaia-kepler-tess-host_2023}) as well as linking planet occurrence and Galactic position or kinematics \citep{winter_stellar_2020, dai_planet_2021, chen_planets_2021-1, longmore_impact_2021, bashi_quantifying_2021, zink_scaling_2023}. It is as-yet uncertain whether correlations between planet occurrence and Galactic position are due to the Galactic environment directly (e.g. \citealt{cai_stability_2017, ndugu_planet_2022}) or reflect a separate, underlying relationship between planet occurrence and stellar metallicity or age (e.g. \citealt{nielsen_planet_2023}).

Investigating how the orbits of planets evolve over time is key to fully understanding how they form, the dependence of planet properties on stellar properties (such as metallicity), and prospects for habitability. However, this investigation likely hinges on our knowledge of stellar age, which is notoriously difficult to constrain. A precise understanding of stellar age is even more elusive for the predominant exoplanet hosts, FGK-type dwarfs and (especially) M dwarfs. Small, rocky, habitable-zone exoplanets are abundant around the lowest-mass stars (e.g. \citealt{mulders_increase_2015}) and relatively easy to observe using transit surveys due to their relatively large planet radius-stellar radius ratio \citep{tarter_reappraisal_2007}. Unfortunately, it is difficult to constrain ages for the lowest-mass stars because their lifetimes are often longer than the age of the universe \citep{choi_mesa_2016}. Isochrone grids are additionally an unreliable source of age measurements, as they often fail to reproduce empirical constraints for M dwarfs due to the presence of strong magnetic fields and starspots \citep{feiden_reevaluating_2012, boyajian_slar_2012, mann_how_2019}.

Though measuring ages for low-mass stars is uniquely challenging, measuring ages for stars of any mass is non-trivial. Stellar ages cannot be directly measured; they can only be inferred using techniques such as comparing measured properties to isochrone models and asteroseismology (e.g. \citealt{karatas_kinematics_2005, lebreton_stellar_2009, silva_aguirre_ages_2015}) or using empirical relations such as gyrochronology and age-activity relations (e.g. \citealt{barnes_ages_2007, soderblom_chromospheric_1991}; for review, \citealt{soderblom_ages_2010}). Recent studies such as \citet{lu_gyro-kinematic_2021} also combine such empirical methods with certain stellar kinematic information. These quantities are associated with relative ages with varying levels of precision (e.g. \citealt{epstein_how_2013, tayar_guide_2022}).
Additionally, the introduction of large-scale spectroscopic surveys including the RAdial Velocity Experiment (RAVE, \citealt{steinmetz_radial_2006}), the Large Sky Area Multi-Object Fiber Spectroscopic Telescope (LAMOST, \citealt{zhao_lamost_2012, cui_large_2012}), the Galactic Archaeology with HERMES survey (GALAH, \citealt{de_silva_galah_2015, martell_galah_2017}), and the Apache Point Observatory Galactic Evolution Experiment (APOGEE, \citealt{majewski_apache_2017}) has delivered spectroscopic data for millions of stars. With these, several groups have recently attempted to obtain spectroscopic stellar ages \citep[e.g.][]{queiroz_starhorse_2018, queiroz_bulge_2020, queiroz_starhorse_2023, mints_unified_2017, mints_isochrone_2018}.

We are motivated by the prospect of conducting large-scale demographic studies on how the orbital dynamics and properties of exoplanets change over time. We now have high-precision astrometric information from \textit{Gaia} homogeneously derived for virtually every nearby star. With stellar parallax, proper motion, and radial velocity measured by \textit{Gaia} along with some assumed potential model of the Milky Way, we can compute a star's Galactic orbit. It has been established for decades that there is a relationship between the velocity dispersion of stars and age (the age--velocity dispersion relation, or AVR) (e.g. \citealt{stromberg_motions_1946, wielen_diffusion_1977}), and dynamical ages have shown potential in calibrating ages even for low-mass dwarf stars \citep[e.g.][]{veyette_chemo-kinematic_2018, newton_new_2018, lu_this_2024, lu_abrupt_2024}. We leverage this nearly stellar-mass-independent Galactic orbit information to construct an age--Galactic orbit relation based on a calibration sample of stars with independently inferred ages.

We note that there is as yet no definite consensus on the dominant cause of the AVR. The AVR may be primarily driven by the orbital heating histories of stars, such that the orbits of older stars have been kinematically heated over time \citep{spitzer_possible_1953}. The effect of disk heating on the AVR has been studied through various N-body and cosmological simulations \citep{hanninen_simulations_2002, holmberg_geneva-copenhagen_2007, house_disc_2011, aumer_agevelocity_2016, grand_vertical_2016}. Alternatively, the AVR may be observed as a consequence of the birthplaces of stars in the galaxy over time. Older stars may have been born kinematically hotter than younger stars, which form with a smaller velocity dispersion at later times, following a so-called ``upside-down" and ``inside out" galactic assembly process \citep{bird_inside_2013, bird_inside_2021}. Indeed, there is evidence to suggest that both mechanisms may contribute significantly to the observed AVR \citep{mccluskey_disc_2024}. We acknowledge that if disk heating is the primary driver of the AVR, then mapping stellar age to galactic orbits will be weakly dependent on stellar mass. Even in this case, dynamically-derived stellar ages will still be far less dependent on stellar mass than e.g. ages from isochrones or asteroseismology. To account for this lack of consensus in the primary cause of the AVR, we call the dynamical age model ``generally mass-independent".

Though we cannot directly measure the ages of stars, these dynamical age estimates provide homogeneous age predictions for hundreds of millions of stars with a wide range of stellar masses and properties. Because these age predictions depend only on dynamical data and the physical assumptions of galactic dynamics, they can be measured independently from other stellar properties such as metallicity and stellar rotation. Such age measurements may enable us to disentangle the relationship between stellar age, metallicity, and other processes of interest (such as exoplanet orbital dynamics).

Due to systematic offsets between different age-dating methods, it is preferable that age predictions for an entire sample of interest are determined using the same method. Most samples of homogeneously inferred stellar ages are small, especially for main-sequence stars; for example, there are only $\mathcal{O}(10^{2})$ main-sequence stars with precise asteroseismic age measurements \citep{huber_fundamental_2013, silva_aguirre_ages_2015}. Most stellar age prediction methods also depend strongly on stellar mass, further limiting the available age samples within a certain stellar mass bin \citep{barnes_ages_2007, soderblom_ages_2010}. This method of deriving dynamical stellar ages is particularly useful because it can be used to generate age predictions for all stars with \textit{Gaia} radial velocities in a homogeneous way for a wide range of stellar masses.

In this work, we introduce a fully dynamical stellar age--vertical action relation calibrated on independently-measured ages of red giant branch and main-sequence turn-off stars. We release the software used in this analysis as an open-source dynamical age prediction tool called \texttt{zoomies}. We validate and compare dynamical age predictions for stars with independent asteroseismic ages, open clusters, and the large stellar samples of the \textit{Kepler}, K2 and \textit{TESS} missions. This paper is organized as follows: in Section \ref{sec:calibration_sample} we describe the stellar samples used in calibration and validation. In Section \ref{sec:methods} we describe our methods, including calculating dynamical properties, hierarchical model construction, and model fitting. In Section \ref{sec:validation}, we discuss internal validation checks on the dynamical age model. In Section \ref{sec:results}, we introduce the model as an open-source age prediction tool and discuss dynamical age predictions for external stellar samples. In Section \ref{sec:discussion}, we discuss how our assumptions may affect our results, and we conclude in Section \ref{sec:conclusions}.

\section{Data} \label{sec:calibration_sample}

\subsection{Calibration Samples} \label{sec:c_sample}

We construct and calibrate two models to predict stellar age from Galactic orbital quantities. We employ two samples of independently measured stellar ages to calibrate the models (the \textit{Calibration Samples}). We use a set of 20,917 red giant branch stars with asteroseismic ages from \citet{stokholm_unified_2023} (hereafter the red-giant calibration sample, or ``RGB" sample). We use only the stars from \citet{stokholm_unified_2023} that have radial velocities measured with \textit{Gaia}. We separately use a set of 33,715 main-sequence turn-off stars from the APOGEE DR17 sample with ages derived with the Bayesian isochrone-fitting code StarHorse \citep{queiroz_starhorse_2023} (hereafter the main-sequence turn off calibration sample, or the ``MSTO" sample). We take only stars labeled as main-sequence turn-off stars by  \citet{queiroz_starhorse_2023} and filter on the included age warning flag ``age\_inout" indicating significant differences between input and output  $\mathrm{log}$ $g$, \teff, and median metallicity. Using these two stellar age calibration samples, we calibrate two distinct dynamical age models; with this, we demonstrate that all dynamical age models depend on the properties of the calibration sample, especially their \lnjzsp distribution (see Figure \ref{fig:RGB_MSTO}).

\subsection{Prediction Sample} \label{sec:p_sample}

\begin{figure*}
    \centering
    \includegraphics[width=\textwidth]{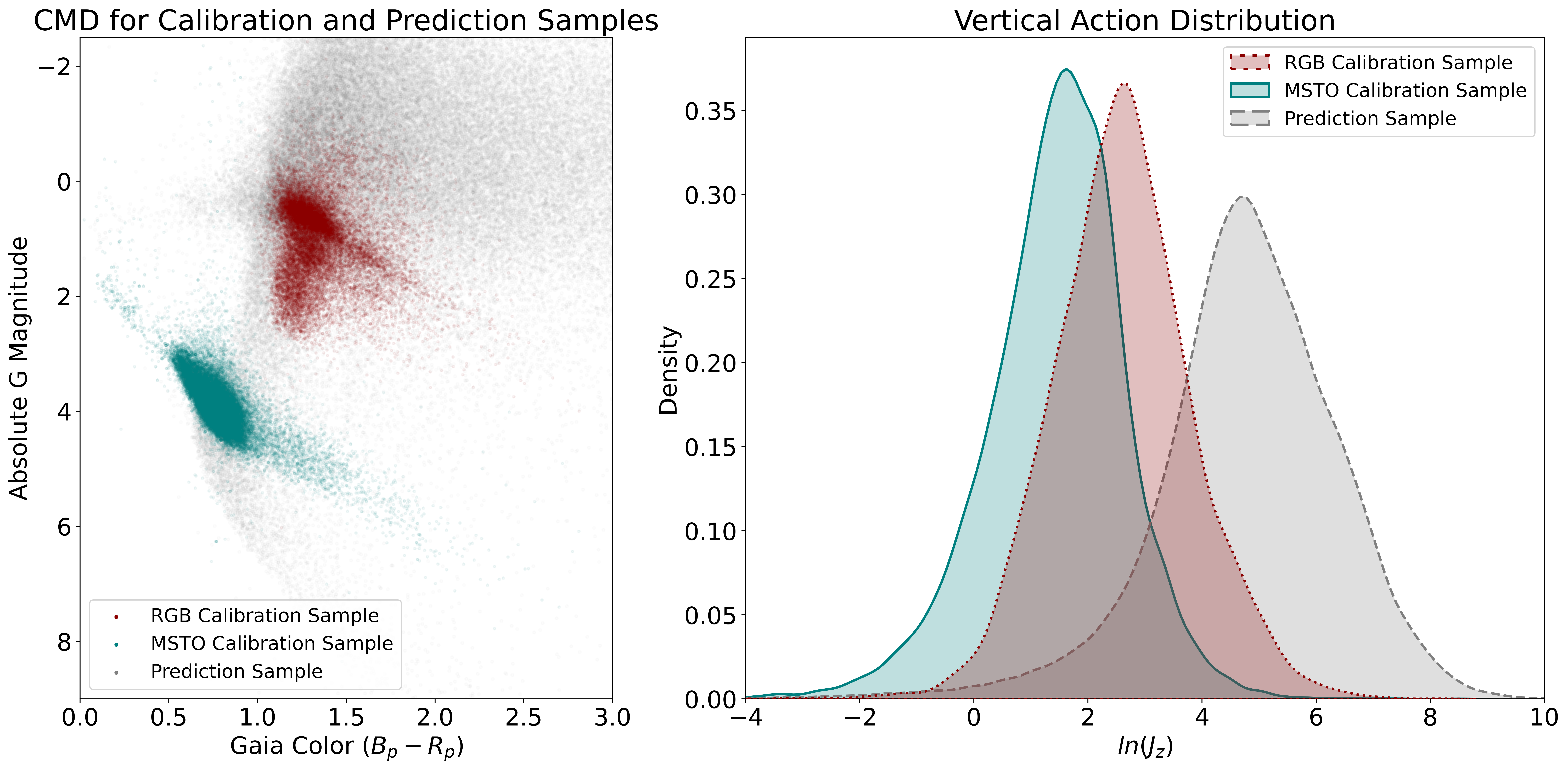}
    \caption{\textit{Left:} Color-magnitude diagram of the RGB calibration sample (red), MSTO calibration sample (blue), and a representative prediction sample (gray). The representative prediction sample consists of the first 100,000 stars returned from a \textit{Gaia} search excluding objects with no radial velocity measurements and $\varpi \leq 0$. \textit{Right:} Vertical action distributions for the RGB (red), MSTO (blue) and representative \textit{Gaia} sample (gray).}
    \label{fig:hrdiagram}
\end{figure*}

The \textit{Prediction Samples} refer to the stellar samples we generate age predictions for. This potentially includes all \textit{Gaia} stars with radial velocity measurements. In this paper, we perform a preliminary analysis using the subset of \textit{Kepler}, K2, and \textit{TESS} main-sequence target stars with \textit{Gaia} DR3 radial velocity measurements. In Figure \ref{fig:hrdiagram}, we present a visual comparison of the RGB Sample, MSTO Sample, and the Prediction Sample including all stars with \textit{Gaia} DR3 radial velocities. In the left panel, we show a color-magnitude diagram (CMD) of the three samples with \textit{Gaia} $B_p-R_p$ color and absolute \textit{Gaia} (G) magnitude calculated from the G-band mean magnitude and the \textit{Gaia} parallax. In the right panel, we compare the vertical action distributions for each of the three samples. 

Whether stellar age estimates are generated with model-dependent, empirical, or kinematically motivated methods, these ages will be dependent on the calibration sample used. Different age predictions will invariably be offset according to the samples used to calibrate the age prediction model. The dynamical age predictions we present are no different; the age--action relations will be best calibrated in the regions of \lnjzsp where the calibration samples lie. Unless the prediction sample of interest is comparable to the calibration sample in \lnjzsp space, dynamical ages should generally be treated as relative, rather than absolute, age predictions. We show in the right panel of Figure \ref{fig:hrdiagram} that the \lnjzsp distribution of the least restrictive prediction sample (all stars with \textit{Gaia} radial velocities) differs significantly from the \lnjzsp distribution of both the RGB and MSTO samples; in such cases, we warn that stellar ages should be interpreted more conservatively as relative ages. Even so, understanding relative stellar ages is useful in many contexts (e.g. for investigating the evolution of planetary systems over time), especially since dynamical ages are measured independently of most stellar properties and are available for very large samples of stars. We include more detailed discussions of this effect in Section \ref{sec:MainSamples}.

\section{Methods} \label{sec:methods}

We construct a hierarchical Bayesian model to predict stellar age from Galactic orbital quantities. We first determine which Galactic orbital quantities to use. The ideal orbital quantity should be one that can be calculated homogeneously for a wide range of stars (given an assumed Milky Way potential model), and one that strongly correlates with stellar age. We could construct an age--vertical velocity ($v_z$) relation, but this would introduce a dependence on the temporal position of each star throughout its orbit. Any stars with wide galactic orbits in the $z$-direction that happen to be near the edge of their orbits would have a vertical velocity close to zero, which may unfairly bias their dynamical age prediction. 

One alternative metric is an integral of motion that encodes time-independent information about a star's galactic orbit. Given a static potential model of the galaxy, the action--angle variables are integrals of motion that describe an object's orbit. Instead of vertical velocity, we elect to use an integral of motion that describes the extent of a star's orbit independently from time: the vertical action, \Jz. \Jzsp is the integral of the vertical galactic position $z$ vs. the vertical velocity $v_z$. \Jzsp labels a star's galactic orbit, no matter where a star happens to be in its orbit at the current moment. The vertical velocity $v_z$ varies with time on the orbit, where $J_z$ (along with $z_{max}$ and other orbital constants of motion) would not. The vertical action is defined as

\begin{equation}
    J_z = \frac{1}{2\pi} \oint{\mathrm{d}z\,v_z }
\end{equation}
where $z$ describes the vertical position and $v_z$ describes the vertical velocity of the object, and the integral is over the path of one full vertical orbit. Since \Jzsp is independent of time, we contend that using \Jzsp will produce a tighter age-galactic motion relation than using the vertical velocity alone. However, the $J_z$ computations depend on the assumed galactic potential model and quality of the \textit{Gaia} kinematic data in hand. We discuss the implications of considering only \Jzsp in the age--action model (and not $J_{r}$, the radial action, or $J_{\phi}$, the azimuthal action) in Section \ref{subsec:allactions}.

\subsection{Calibrating the age--action relation}\label{subsec:calibratingageaction}

In this section, we describe the process of constructing and calibrating the age--action relation. We use an axissymetric potential model of the Milky Way to compute the actions, implemented in the Python package \texttt{gala} \citep{price-whelan_gala_2017, price-whelan_adrngala_2020}. This potential model includes a spherical galactic nucleus and bulge, a 3-component sum of Miyamoto-Nagai galactic disks \citep{miyamoto_three-dimensional_1975} representing an approximation to the potential from an exponential disk \citep{smith_simple_2015}, and a spherical Navarro-Frenk-White dark matter halo model \citep{navarro_structure_1996}. The disk model follows the galactic rotation curve from \citet{eilers_circular_2019}, and the vertical structure is set by the shape of the phase-space spiral in the solar neighborhood from \citet{darragh-ford_concentrationmass_2023}. We describe a preliminary test on the impact of potential model choice on final stellar age predictions in Appendix \ref{appendix:galmodels}.

When calibrating the age--\lnjzsp model, we assume that the dynamical ages of stars are independent from stellar mass. We return to this assumption and discuss its implications in Section \ref{subsec:stellarmassdependence}. For each star in each calibration sample, we first calculate the logarithm of the vertical action \lnjzsp given this Milky Way potential model. We calculate \lnjzsp from the \textit{Gaia} proper motion, radial velocity, and using inverse parallax as the distance estimate. We use the \texttt{gala} software to make these calculations. We take the initial conditions given by the \textit{Gaia} data and integrate each star's orbit through time within the galactic potential model. We numerically integrate the orbits to calculate actions using the Staeckel Fudge implementation in the \texttt{Agama} software \citep{vasiliev_agama_2019}. In both calibration samples, the errors in \lnjzsp are much smaller than the intrinsic variance in \lnjzsp (the average \lnjzsp uncertainty in both the RGB and MSTO samples is approximately 0.05, compared to a mean intrinsic variance of approximately 1.5), so we neglect the errors in \lnjz. We performed a preliminary check on whether this intrinsic variance is correlated with any of several stellar parameters including stellar metallicity; $log(g)$; $T_{\mathrm{eff}}$; distance; stellar mass, radius, and luminosity; $\alpha$ abundance; and $\nu_{\mathrm{max}}$. We found no evidence for correlations between the intrinsic variance and any of these stellar parameters. We provide further details of this check in Appendix \ref{appendix:intrinsicscattercheck}.

Next, we construct a hierarchical Bayesian model to describe the age--\lnjzsp relation. In this model, we wish to describe the log of the intrinsic variance (the squared uncertainty) in the age direction, $\mathrm{ln} V_{\tau}$, as a free parameter. Though we intend to construct a hierarchical model to predict stellar age given \lnjzsp (in other words, constructing a $P(\tau | \lnjz)$ model), fitting a model to the age--\lnjzsp relation proves difficult when incorporating $\mathrm{ln} V_{\tau}$. Because the age--\lnjzsp relation is steep and has large scatter in the age direction, the value of the $ \mathrm{ ln} V_{\tau}$ free parameter balloons and fails to converge. To counteract this, we instead consider the \lnjzsp vs. age relation, $P(\lnjz | \tau)$, where $\mathrm{ln} V_{\mathrm{ln}J_{z}}$ describes the intrinsic variance in the \lnjzsp direction. Once the $P(\lnjz | \tau)$ model is calibrated, we obtain  $P(\tau | \lnjz)$ using a process we describe in detail in Section \ref{subsec:stellaragepred}.

Since we calibrate the model on samples of main-sequence turn-off stars and red giant branch stars, the number density of stars along the age axis will reflect these samples. For example, the MSTO stellar sample has a high number density near 10 Gyr (see Figure \ref{fig:RGB_MSTO}, bottom left panel), which would unfairly bias the age prediction towards 10 Gyr when applied to other populations of stars with different number density distributions, such as main-sequence FGKM dwarfs. To counteract this effect, we estimate the stellar number density along the age axis $\lambda(\tau)$ as part of the age--action model, and we marginalize out $\lambda(\tau)$ when applying the model to other stars. This process enables us to apply only the slope and variance of the age--action model to other stellar samples, without also applying the calibration sample's number density distribution. Accounting for the number density in this way produces a more generally applicable dynamical age prediction model.

Our method will not only require calibrating the age--action relationship, but also the calibration sample's stellar number density in age, $\lambda(\tau)$. To estimate this, we include a Poisson likelihood component in the model likelihood function, similar in structure to the one used by \citet{everall_photo-astrometric_2022} in a different context. If the number density profile of stars along the age axis is $\lambda(\tau)$, we take the probability of drawing a population of stellar ages $\tau_i$ from $\lambda(\tau)$ to be the Poisson likelihood function (derived in \citealt{lombardi_fitting_2013, everall_seestar_2020}):

\begin{equation} \label{eq:poisson}
    \ln \mathcal{L_{\mathrm{Poisson}}} = \sum_{i=1}^{N} \ln \lambda(\tau_i) - \int{ \lambda(\tau) \mathrm{d}\tau}
\end{equation}
In the hierarchical model, we describe the number density $\mathrm{ln}\lambda(\tau)$ as a third-degree spline defined between 0 and 14 Gyr. We add the Poisson likelihood factor into the complete model likelihood function. By independently fitting $\lambda(\tau)$ of the calibration sample in the age--action model, all other model parameters describing the age--\lnjzsp relation are not significantly influenced by the calibration sample's non-uniform $\lambda(\tau)$. The $\lambda(\tau)$ distribution is specific to each calibration sample and is dependent on stellar mass, selection functions, and any number of other factors; we do not want the $\lambda(\tau)$ distribution of the calibration sample to affect the prediction sample's dynamical age estimates. By fitting the calibration sample's $\lambda(\tau)$ distribution and ignoring it when using the age--\lnjzsp model to generate age predictions for other stars, we minimize the risk of the calibration sample's unique $\lambda(\tau)$ distribution affecting dynamical age estimates for other stars. 

Finally, we model the \lnjzsp--age slope as a monotonic quadratic spline (a quadratic spline that must increase at each successive knot value). 

\begin{figure*}
    \centering
    \includegraphics[width=\linewidth]{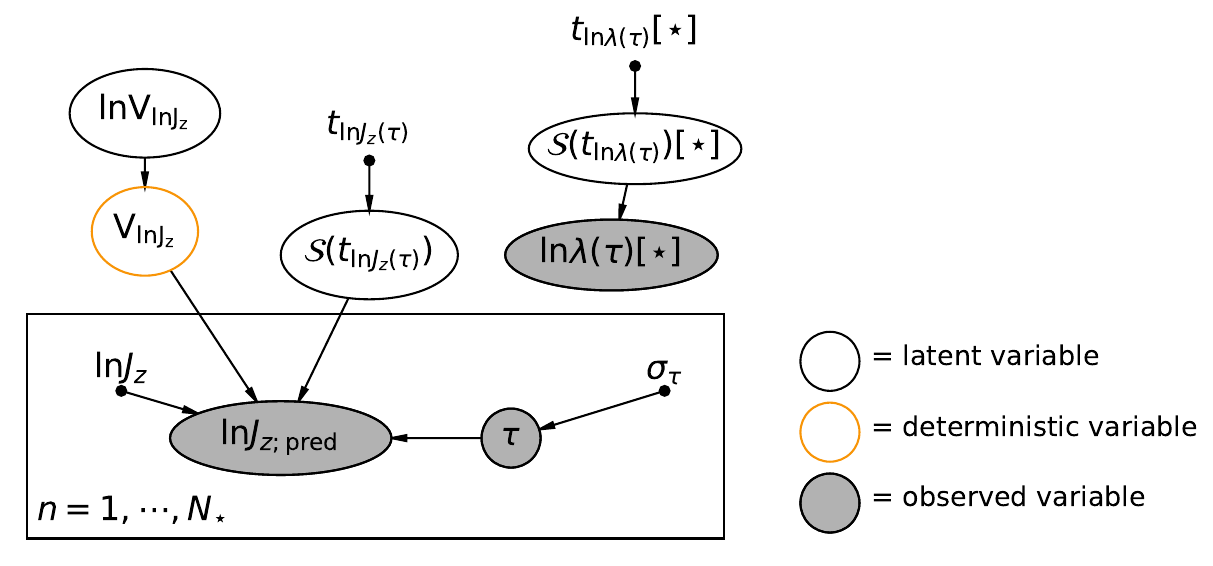}
    \caption{Graphical model representation of the hierarchical Bayesian stellar age--\lnjzsp model. $t_{\tau(\mathrm{ln}J_z)}$ and $t_{\mathrm{ln}\lambda(\tau)}$ refer to the fixed spline knot vectors for the $\tau(\mathrm{ln}J_z)$ and $\mathrm{ln}\lambda(\tau)$ splines, respectively, and $\mathcal{S}(t_{\tau(\mathrm{ln}J_z)})$ and $\mathcal{S}(t_{\mathrm{ln}\lambda(\tau)})$ represent the associated knot value free parameters. A list of the model parameters and their priors is given in Table \ref{table:ModelParams}. As in Table \ref{table:ModelParams}, ``$[\star]$" denotes a stellar number density model parameter that we ignore when we apply the calibrated model to a prediction sample of stars.}
    \label{fig:graphical_model}
\end{figure*}

We summarize the model fixed and free parameters here. The knot locations of the $\tau(\mathrm{ln}J_z)$ monotonic spline, $ t_{\tau(\mathrm{ln}J_z)} $, are fixed at every $3.75$ Gyr between ${-1}$ and $14$ Gyr. The knot locations of the $\mathrm{ln }\lambda(\tau)$ third-degree spline, $t_{\mathrm{ln}\lambda(\tau)}$, are fixed at every $1$ Gyr between ${-1}$ and $14$ Gyr. The model free parameters include the knot values of the $\tau(\mathrm{ln}J_z)$ monotonic spline $\mathcal{S}(t_{\tau(\mathrm{ln}J_z)})$, the knot values of the $\mathrm{ln }\lambda(\tau)$ third-degree spline $\mathcal{S}(t_{\mathrm{ln}\lambda(\tau)})$, and the intrinsic variance $\mathrm{ln} V_{\mathrm{ln}J_{z}}$. We show the graphical model representation of this model in Figure \ref{fig:graphical_model}. A list of the model free and fixed parameters is provided in Table \ref{table:ModelParams}. The model parameters associated with $\mathrm{ln} \lambda(\tau)$ which are ignored when using the model to generate age predictions for other stars are marked with ``$[\star]$".

\begin{deluxetable*}{c|c|c}
\tablecaption{Dynamical age--action model parameters \label{table:ModelParams}}
\tablehead{\colhead{Parameter} & \colhead{Prior} & \colhead{Description}}
\startdata
$ \mathrm{ln}{V_{\mathrm{ln}J_z}} $ & $\mathcal{N}(9,5)$ & Log of intrinsic variance \\
$ V_{\mathrm{ln}J_z} $ & Deterministic & Intrinsic variance \\
$ t_{\tau(\mathrm{ln}J_z)} $ & Fixed & Knot vector for $\tau(\mathrm{ln}J_z)$ spline, where $t_{\tau(\mathrm{ln}J_z)} = (t_{\tau(\mathrm{ln}J_z),1} ... t_{\tau(\mathrm{ln}J_z),5})$\\
$ \mathcal{S}(t_{\tau(\mathrm{ln}J_z)}) $ & $\mathcal{U}(0,5)$ & Knot values for $\tau(\mathrm{ln}J_z)$ spline, defined at each point in the knot vector  $t_{\tau(\mathrm{ln}J_z)}$ \\
$ t_{\mathrm{ln}\lambda(\tau)} [\star] $ & Fixed & Knot vector for $\mathrm{ln}\lambda(\tau)$ spline, where $t_{\mathrm{ln}\lambda(\tau)} = (t_{\mathrm{ln}\lambda(\tau),1} ... t_{\mathrm{ln}\lambda(\tau),15})$\\
$ \mathcal{S}(t_{\mathrm{ln}\lambda(\tau)}) [\star] $ & $\mathcal{U}(-10,15)$ & Knot values for $\mathrm{ln}\lambda(\tau)$ spline, defined at each point in the knot vector  $ t_{\mathrm{ln}\lambda(\tau)} $\\
$ ln \lambda(\tau) [\star] $ & Observed & Log of stellar number density \\
$ \tau $ & Observed & Stellar age \\
$ \mathrm{ln}J_{z, pred} $ & Observed & Predicted \lnjzsp \\
\enddata
\tablecomments{The priors listed for the spline knot values $\mathcal{S}(t_{\tau(\mathrm{ln}J_z)})$ and $\mathcal{S}(t_{\mathrm{ln}\lambda(\tau)})$ apply independently for each knot. A graphical model representation of the model is shown in Figure \ref{fig:graphical_model}.}
\tablecomments{ $[\star]$: We fit $\lambda(\tau)$ in the dynamical age model, but we do not apply $\lambda(\tau)$ to generate dynamical age predictions for the prediction sample. We ignore these parameters when we apply the calibrated model to a prediction sample of stars.}
\end{deluxetable*}

We generate two mean posterior age--Jz relations: one using the red-giant branch star calibration age sample and one using the main-sequence turn off star calibration age sample. For each calibration sample, we sampled the free parameters using No U-Turn Sampling \citep{homan_no-u-turn_2014} with the Python package numpyro \citep{phan_composable_2019, bingham_pyro_2019} with chains of 2,000 draws each. We use 2,000 tuning steps for each fit. To check for convergence, we calculate the Gelman-Rubin $\hat{r}$ summary statistic for each model free parameter \citep{gelman_inference_1992}. The $\hat{r}$ statistic checks for convergence by comparing the level of variation among multiple MCMC chains. If the sampler successfully converged, $\hat{r}$ is expected to be within $0.1$ of $1$. For both the RGB and MSTO models, all free parameters converged with $1.00 < \hat{r} < 1.05$.

\subsection{Generating stellar age predictions} \label{subsec:stellaragepred}

\begin{figure}
    \centering
    \includegraphics[width=\linewidth]{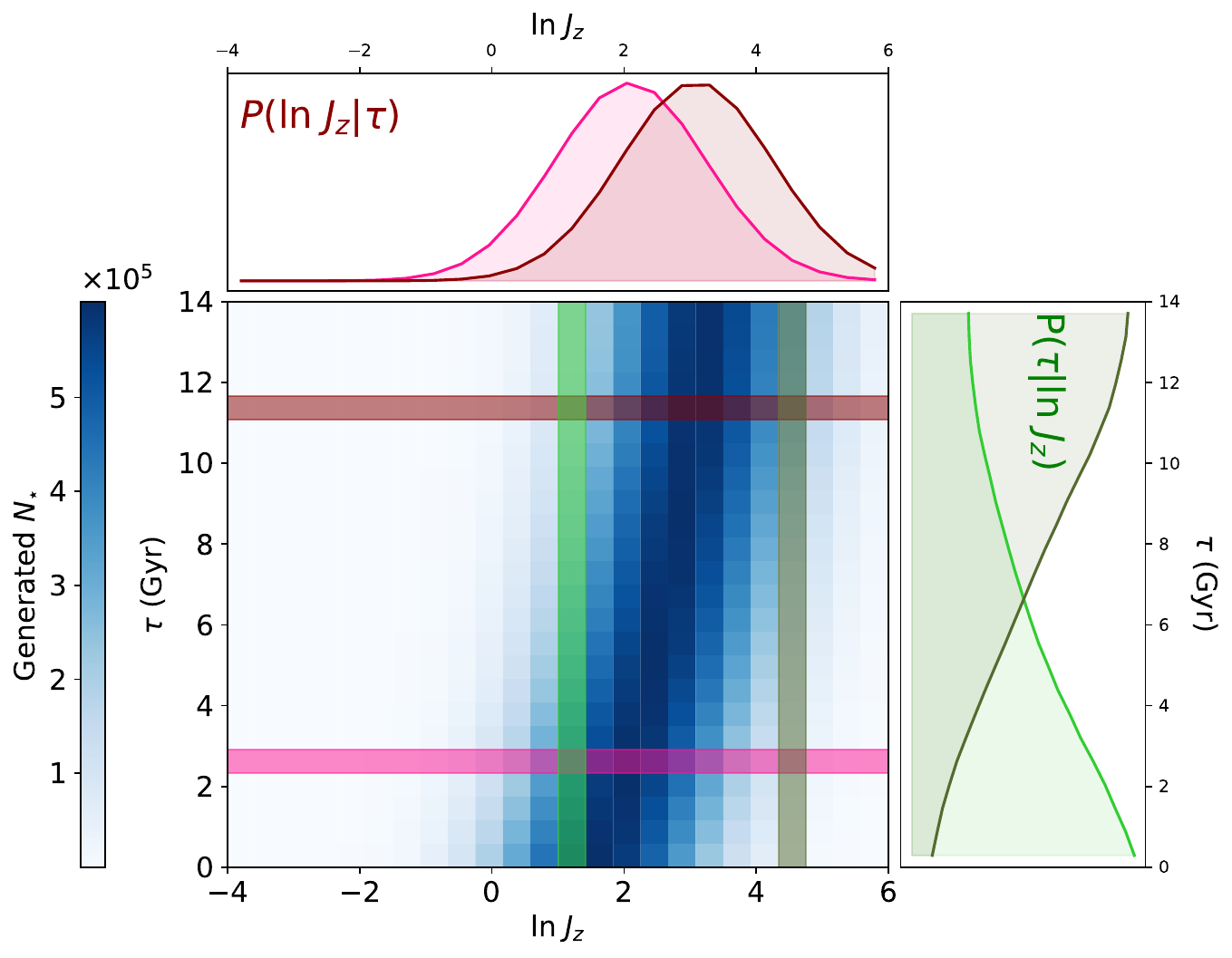}
    \caption{Illustration of the conditional probabilities $P(\lnjz | \tau)$ and $P(\tau | \lnjz )$ in relation to the joint probability $P(\tau, \lnjz)$. The blue heatmap represents the joint probability $P(\tau, \lnjz)$ and shows a simulated stellar sample in $\tau$--\lnjzsp space generated from the mean posterior dynamical age model. The pink and red highlighted horizontal regions correspond to the conditional probability $P(\lnjz | \tau)$, shown on the top panel as probability density functions in the \lnjzsp dimension. The light and dark green highlighted vertical regions correspond to the conditional probability $P(\tau | \lnjz)$, shown on the right panel as probability density functions in the $\tau$ dimension. The $P(\tau | \lnjz)$ probability density functions are analogous to stellar age predictions based on \lnjzsp from \textit{Gaia}.}
    \label{fig:conditional_prob}
\end{figure}

Given a value of \lnjz, we use the calibrated age--vertical action relation to predict an age probability distribution. When using the mean posterior age--\lnjzsp model to predict ages for stars with \lnjz, we ignore the number density in age. In other words, we consider only the mean and intrinsic variance of the age--\lnjzsp relation, and we do not consider the number density in age of the calibration samples when generating stellar age predictions.

Due to the calibration samples' large scatter on the age axis, we construct the age-action model to find $P(\lnjz | \tau)$, as described in Section \ref{subsec:calibratingageaction}. In order to use this model to predict age from \lnjz, we adjust this model to find $P(\tau | \lnjz)$. We invoke Bayes' theorem:

\begin{equation}\label{eq:bayesjz}
    P(\tau | \lnjz) = \frac{P(\lnjz | \tau) P(\tau)}{P(\lnjz)}
\end{equation}

Because we have taken the step of independently fitting $\lambda(\tau)$ of the calibration sample, the rest of the model parameters described in Section \ref{subsec:calibratingageaction} (namely, $\mathcal{S}(t_{\mathrm{ln}J_z(\tau})$ and $\mathrm{ln}V_{\mathrm{ln} J_z}$) must be sufficiently independent from the stellar number density of the calibration sample. Therefore, we take the stellar age prior $P(\tau)$ to be uniform here. Marginalizing over $\tau$ allows us to calculate the marginal probability $P(\lnjz)$:

\begin{equation} \label{eq:marginalize}
    P(\lnjz) = \int P(\lnjz | \tau) P(\tau) \mathrm{d}\tau
\end{equation}
In Figure \ref{fig:conditional_prob}, we show a visual illustration of the conditional probabilities $P(\tau | \lnjz)$ and $P(\lnjz | \tau)$ in terms of the joint probability $P(\tau, \lnjz)$. We illustrate using a simulated stellar sample generated from the mean posterior dynamical age model with the stellar number density $\lambda(\tau)$ marginalized out.

\begin{figure*}
    \centering
    \includegraphics[width=\textwidth]{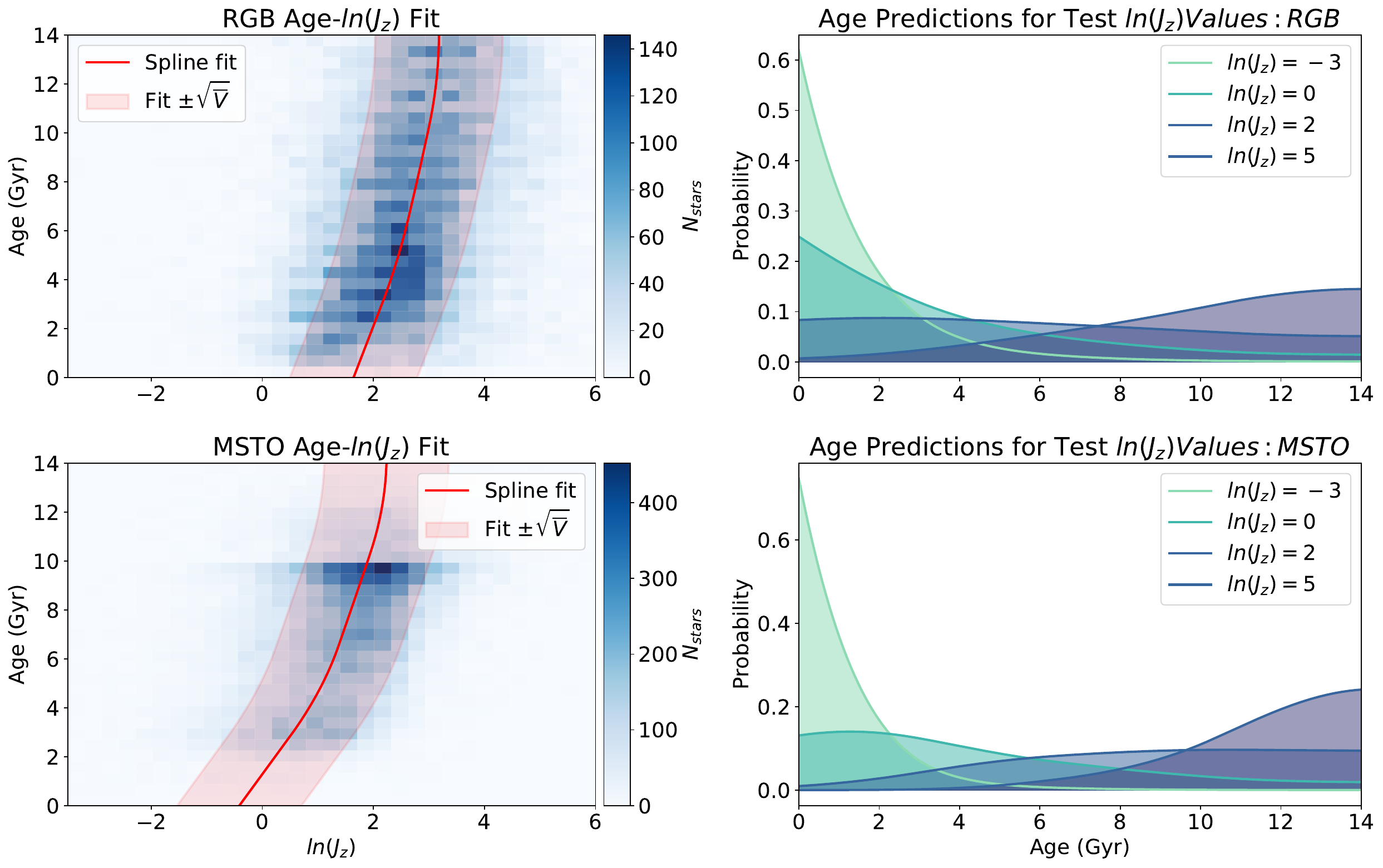}
    \caption{\textit{Top left:} 2-D histogram of the RGB calibration sample in \citet{stokholm_unified_2023} asteroseismic age vs. \lnjzsp space. The mean posterior $\tau$--$\lnjz$ monotonic spline is overplotted, $\pm$ the square root of the best-fit intrinsic variance. \textit{Bottom left:} 2-D histogram of the MSTO calibration sample in \citet{queiroz_starhorse_2023} isochrone ages vs. \lnjzsp space. \textit{Top right:} Generated age prediction distributions for a set of test \lnjzsp values using the RGB-calibrated model. \textit{Bottom right:} Generated age prediction distributions for the same set of test \lnjzsp values using the MSTO-calibrated model.}
    \label{fig:RGB_MSTO}
\end{figure*}

To draw the mean posterior dynamical age model in age--\lnjzsp space, we first generate a test grid \tgridsp of possible stellar ages from 0 to 14 Gyr, arbitrarily spaced by 0.014 Gyr. We then evaluate the $\tau(\lnjz)$ monotonic spline at each point on \tgridsp using the mean posterior $\mathcal{S}(t_{\tau(\mathrm{ln}J_z)})$ knot values. We take the \lnjzsp values on the mean posterior spline corresponding to \tgridsp to be \lnjzgrid. The mean posterior age--\lnjzsp splines evaluated over \tgridsp (in other words, \tgridsp vs. \lnjzgrid) are shown in red in the left panels of Figure \ref{fig:RGB_MSTO}.

To calculate $P(\tau | \lnjz)$ for a star of interest with a given \lnjz, we evaluate Equations \ref{eq:bayesjz} and \ref{eq:marginalize}. We evaluate the integral in Equation \ref{eq:marginalize} numerically with Simpson's rule across \lnjzgridsp to generate a stellar age probability distribution across \tgrid. With this, we are able to take any given \Jzsp value and generate a predicted dynamical age probability distribution, evaluated across \tgrid.

In Figure~\ref{fig:RGB_MSTO}, we present a comparison of the age--\lnjz\ model fit between the RGB calibration sample and MSTO sample. On the left panels, we show both calibration samples in (asteroseismic or isochrone) age vs. \Jzsp space with the best-fit model over-plotted. On the right panels, we show generated age prediction distributions for a set of test \lnjzsp values for both models. We demonstrate that the analogous test age predictions differ, especially for large \Jz, due to the \Jzsp distributions of the calibration sample. However, both models agree in terms of relative age (i.e. when stars are ranked in order of youngest to oldest).

\section{Validation} \label{sec:validation}

\subsection{Internal Model Validation}

\begin{figure*}
    \centering
    \includegraphics[width=\textwidth]{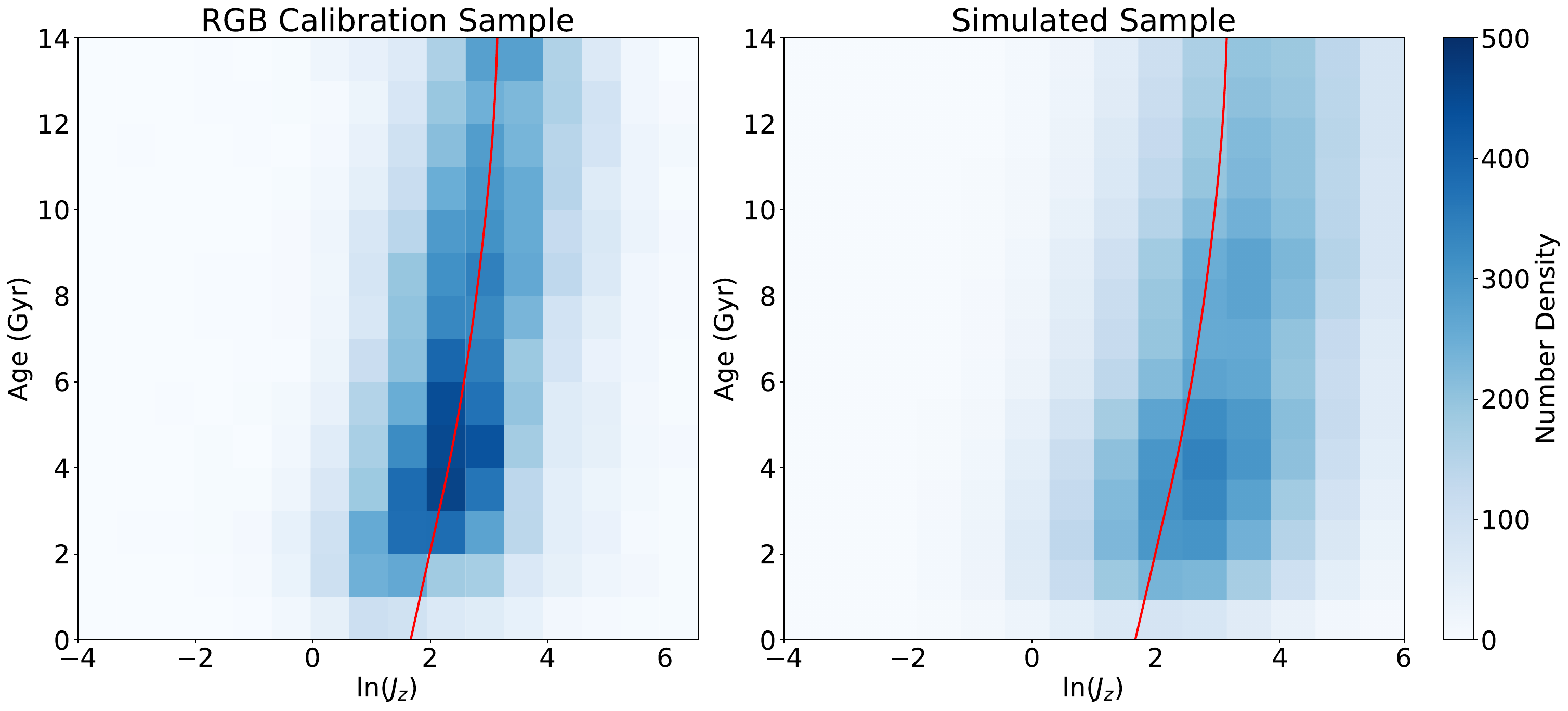}
    \caption{\textit{Left:} 2-D histogram of the red giant calibration sample, and the mean (ridgeline) of the model fit (red). \textit{Right:} Predicted density of stellar ages from model fit (red). While we fit and can re-generate the stellar number density along the age axis according to the calibration sample, we ignore this number density when predicting stellar ages, effectively using only the age--\Jzsp spline model represented in red.}
    \label{fig:validation}
\end{figure*}

We internally validate the model using a $k$-fold cross-validation method. We split each calibration sample randomly into 10 equal subsamples. We designate one subsample as the ``withheld" subsample, and the other nine subsamples combined as the ``validation" sample. We calibrate the dynamical age model using the validation sample. We then evaluate the calibrated model's goodness of fit using the withheld subsample, with

\begin{equation}
    \chi = \frac{\tau - \tau_{pred}}{\sqrt{V + \sigma_{\tau}^2}}
\end{equation}
where $\tau$ represents the stellar ages for the withheld sample, $\tau_{pred}$ represents the mean dynamical age predictions for the withheld sample, V is the mean posterior intrinsic variance, and $\sigma_{\tau}$ represents the stellar age error for the withheld sample. We repeat this process 10 times, where each of the 10 subsamples in turn becomes the withheld sample. We verify that $\sigma_{\chi}$ (the standard deviation of $\chi$ for each of the ten iterations) is close to 1.

We additionally verify that from the calibrated dynamical age model, we are able to reproduce both the age--action trend and the stellar number density of the calibration sample. A visualization of this verification is shown in Figure \ref{fig:validation}. On the left panel, we show  \lnjzsp vs. age for the RGB calibration sample as a 2-D histogram with arbitrary bins, along with the best-fit age--\Jzsp model shown in red. The color corresponds to the number density of stars appearing in each bin. On the right panel, we show the number density in each bin predicted by our model. The dynamical age model encodes information about both the age--action trend and the calibration sample number density; however, in predicting ages for stars given \lnjz, we do not consider the number density factor $\lambda(\tau)$.

\subsection{Comparison with Asteroseismic RGB Ages}

\begin{figure}[ht!]
    \centering
    \includegraphics[width=\linewidth]{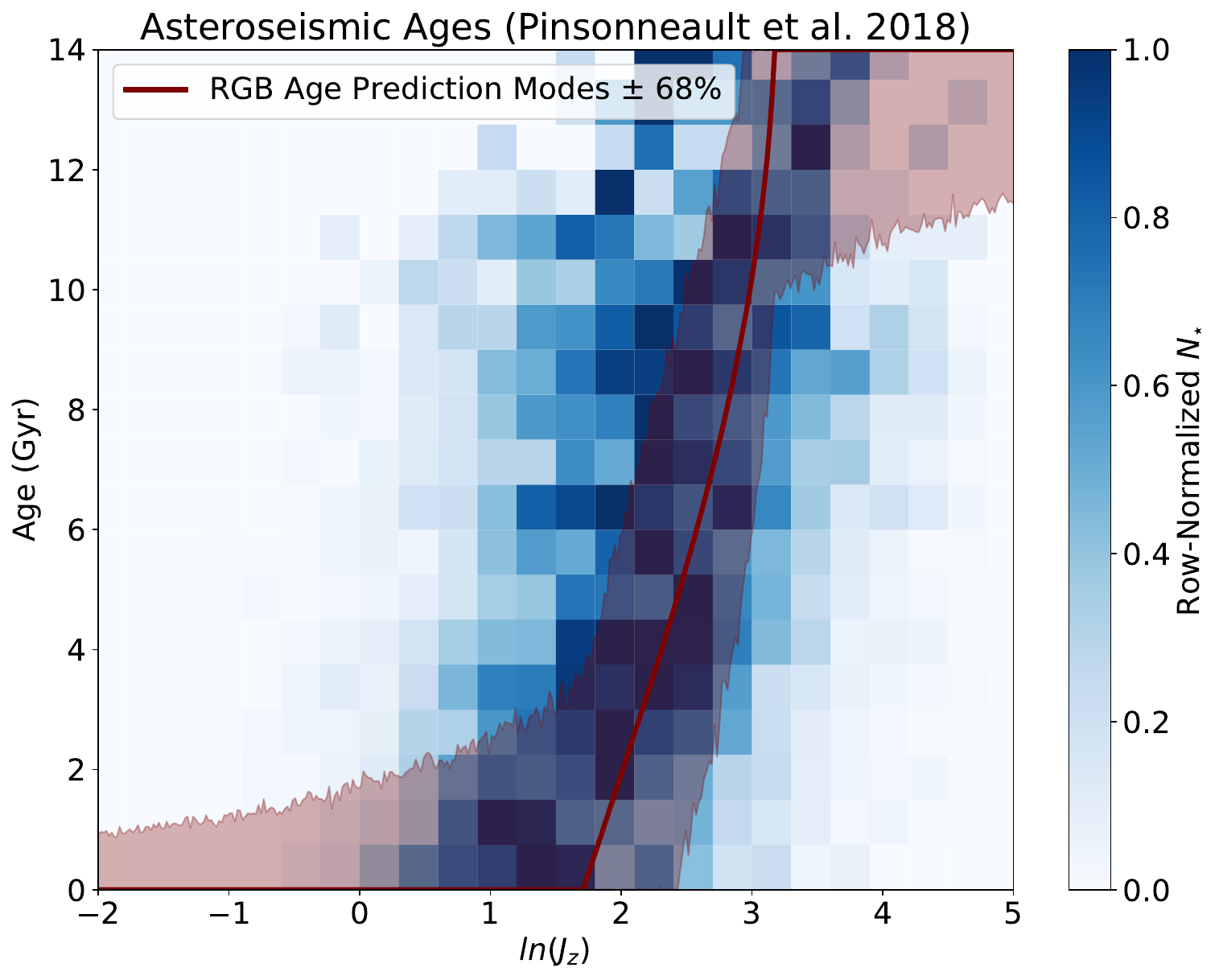}
    \caption{A 2-D histogram of the \citet{pinsonneault_second_2018} sample of red-giant-branch stars in asteroseismic age-\lnjzsp space. The RGB-calibrated dynamical ages for this sample are designated with the red line and shaded region. The solid red line represents the statistical mode of each dynamical age probability distribution associated with each star in the \citet{pinsonneault_second_2018} sample, and the shaded region represents $68\%$ of the distribution mass around the mode. The 2-D histogram is row-normalized to reduce the visual effect of the stellar density distribution on the $\tau$ axis.}
    \label{fig:asteroages}
\end{figure}

We validate our model-predicted ages against another independent sample of red giant branch stars with asteroseismic age predictions. In Figure \ref{fig:asteroages}, we show a column-normalized heatmap of stellar ages from the APOKASC catalog \citep{pinsonneault_second_2018}, which derived spectroscopic and asteroseismic ages for red giant stars. The \citet{pinsonneault_second_2018} sample closely matches the RGB calibration sample in terms of stellar mass and temperature distribution, but have ages determined independently of dynamics. We calculated vertical actions for the stars in the \citet{pinsonneault_second_2018} sample that have \textit{Gaia} DR3 radial velocities. With Figure \ref{fig:asteroages}, we demonstrate that our dynamical age predictions are consistent in age--\lnjzsp space with the \citet{pinsonneault_second_2018} sample, where the \lnjzsp distribution is comparable to our calibration sample. The age predictions based on dynamics alone are both truncated and poorly constrained, but we seek to demonstrate that our dynamical age predictions are consistent with independent age measurements.

\section{Results} \label{sec:results}

We publish the dynamical age model as an open-source, user-friendly tool called \texttt{zoomies}\footnote{www.github.com/ssagear/zoomies} \citep{sagear_ssagearzoomies_2024}. This tool takes an input of \textit{Gaia} data for any star with a \textit{Gaia} radial velocity and outputs a predicted age distribution. We use this tool to generate dynamical age predictions for several samples of stars.

We note that dynamical age predictions generated by this tool are dependent on the precision and accuracy of ages in the calibration samples, as seen in the differing age probability distributions given \lnjzsp associated with each calibrated model, shown in Figure \ref{fig:RGB_MSTO}. Indeed, the uncertainty in the age-\lnjzsp relation contributes significantly to the wide tails of the age probability distributions (the median age uncertainty for the RGB sample from \citet{stokholm_unified_2023} $\sigma_{\tau} \approx 1$ $\mathrm{Gyr}$, and for the MSTO sample from \citet{queiroz_starhorse_2023} $\sigma_{\tau} \approx 0.5$ $\mathrm{Gyr}$). Age estimates generated for stellar populations that are similar to the calibration sample (in terms of $M_{\star}$ and \lnjz) may be considered in an absolute sense. However, for stellar populations that differ significantly from the calibration sample, it may be more useful to consider the generated age estimates as relative ages, e.g. in comparing older vs. younger groups within a larger population. We perform a preliminary test to check this assertion, described in Appendix \ref{appendix:relative}.

\texttt{zoomies} offers functionality for users to calibrate a dynamical age model using any calibration samples with externally measured ages and \textit{Gaia} astrometric data; the precision of the resulting age estimates will depend on the age uncertainties in the chosen calibration sample and the associated scatter along the \lnjzsp axis. In the following sections, we compare and discuss our dynamical age estimates for external stellar samples.

\subsection{Comparison with Open Cluster ages} \label{subsec:openclusterages}

\begin{figure*}
    \centering
    \includegraphics[width=\linewidth]{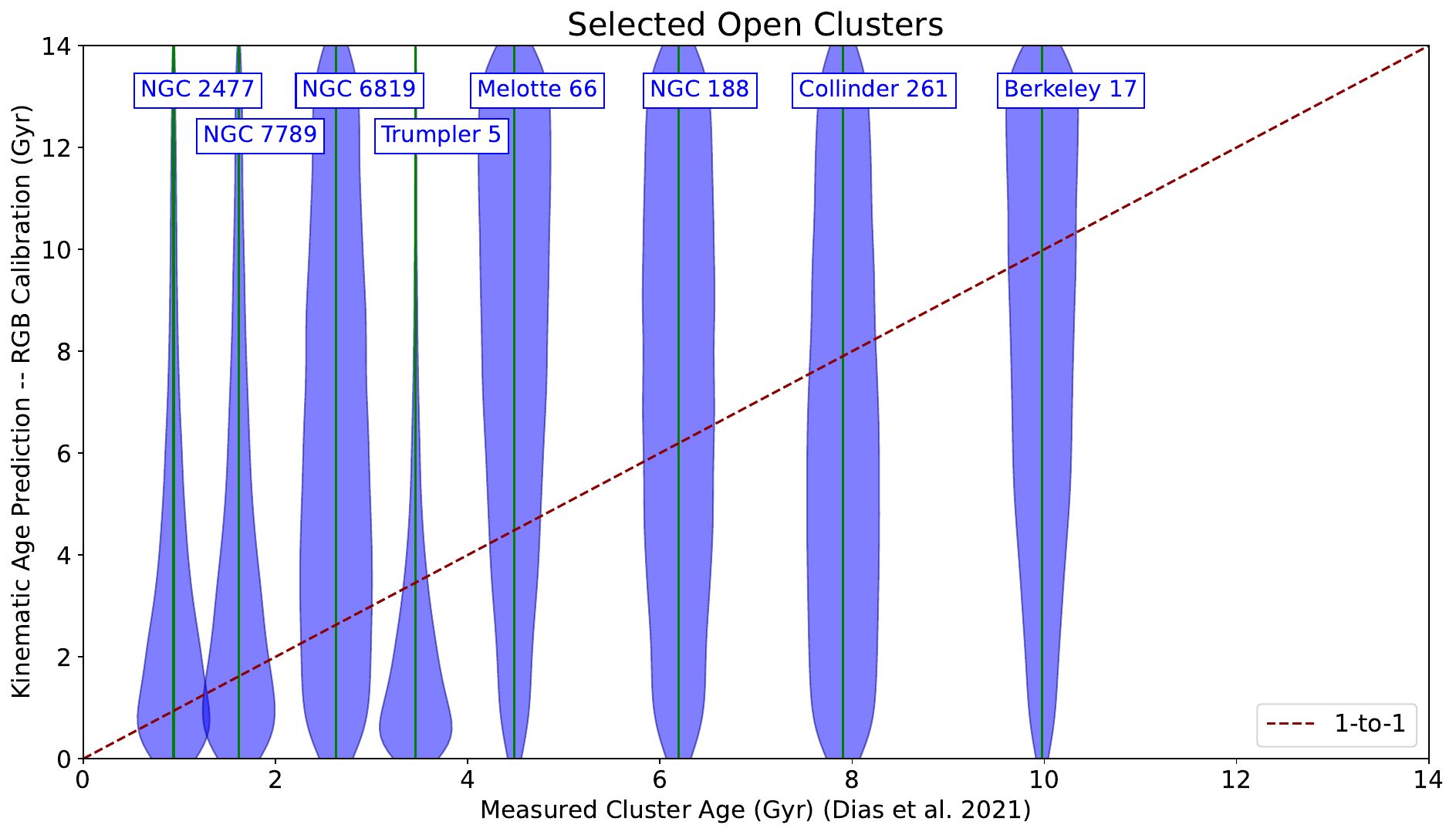}
    \caption{Violin plot of selected open cluster dynamical ages. Each cluster segment is labelled according to the open cluster it represents. Each segment represents the age probability distribution for the median \lnjzsp value for cluster members (with arbitrary width), according to the RGB-calibrated dynamical age model. The position of each cluster segment on the x-axis represents the cluster age reported by \citet{dias_updated_2021}. The one-to-one relation is shown with the red dotted line. We show that there is broad consistency between the cluster-derived ages and dynamical age predictions, though dynamical age constraints are calibration-limited and much less constraining than the cluster age measurements.}
    \label{fig:cluster_plot}
\end{figure*}

We generate dynamical age predictions for open cluster members with independently measured cluster ages. We use the cluster membership lists and ages presented by \citet{dias_updated_2021} to identify individual members of each cluster. We select eight open clusters to present in this paper, and we include an analysis of a random sample of the clusters in \citet{dias_updated_2021} in Appendix \ref{appendix:cluster}.

We select the eight clusters in Figure \ref{fig:cluster_plot} with the following method: first, we bin the cluster ages determined by \citet{dias_updated_2021} from 0 to 14 Gyr with an arbitrary width of 1 Gyr. We do this to select an evenly distributed sample of young and old open clusters. Next, we take the cluster in each bin that has the most members, ensuring that each cluster has more than 100 members at this stage. For each cluster, we omit the members that have a membership probability less than 0.5. We consider only the remaining cluster members that have \textit{Gaia} parallaxes $> 0$ and radial velocities measured. We present the selected clusters, the number of remaining members and their ages in Table \ref{tab:clusters}. We perform the same analysis on a random sample of clusters and include this analysis in Appendix \ref{appendix:cluster}.

Using the \textit{Gaia} data for this sample, we calculate \lnjzsp for each star according to the method described in Section \ref{sec:methods}. For each cluster, we take the median \lnjzsp across all cluster members and generate the corresponding dynamical age probability distribution. We take this age distribution to be the representative age distribution for the entire cluster.

In Figure \ref{fig:cluster_plot}, we show the age distributions for each selected cluster as a violin plot. Though our dynamical age distributions are not as precise as the measured cluster ages, we demonstrate that the relative age predictions for open clusters scale appropriately using only dynamical data.

\begin{deluxetable}{ccc}
\tablewidth{\linewidth}
\tablecaption{Open cluster sample parameters from \citet{dias_updated_2021}. $N$ is the number of cluster members used in analysis after vetting based on availability of \textit{Gaia} parallax and radial velocity, and membership probability $>= 0.5$ according to \citet{dias_updated_2021}. $\tau_{Cluster}$ is the independently measured cluster age from \citet{dias_updated_2021}. \label{tab:clusters}}
\tablehead{\colhead{Cluster} & \colhead{$N$}& \colhead{$\tau_{Cluster}$} \\
\colhead{} & \colhead{} & \colhead{(Gyr)}}
\startdata
NGC 2477 & 421 & 0.94 \\
NGC 7789 & 639 & 1.62 \\
NGC 6819 & 132 & 2.63 \\
Trumpler 5 & 192 & 3.46 \\
Melotte 66 & 69 & 4.49 \\
NGC 188 & 84 & 6.19 \\
Collinder 261 & 105 & 7.90 \\
Berkeley 17 & 25 & 9.98 \\
\enddata
\end{deluxetable}

\subsection{Kepler, K2, and \textit{TESS} main samples} \label{sec:MainSamples}

\begin{figure*}
    \centering
    \includegraphics[width=\textwidth]{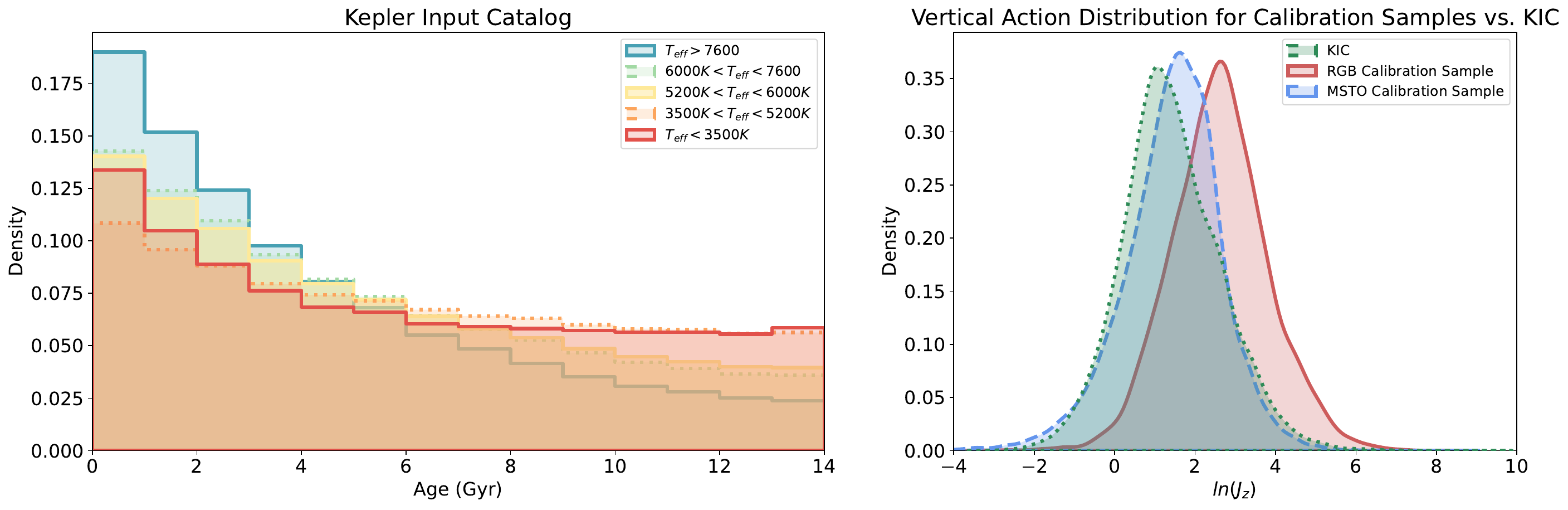}
    \caption{\textit{Left:} Combined dynamical age distributions (calibrated with the MSTO sample) for the KIC sample, partitioned by \teffsp (from \citealt{mathur_revised_2017}). \textit{Right:} \lnjzsp distribution for the entire KIC sample (green) compared to the \lnjzsp distributions for the RGB and MSTO calibration samples (red and blue, respectively).}
    \label{fig:KeplerAges}
\end{figure*}

In this Section, we investigate dynamical ages derived with \texttt{zoomies} for several samples of stars from large exoplanet transit surveys. We employ the \textit{Kepler} Input Catalog (KIC) \citep{brown_kepler_2011} together with the \textit{Gaia}-\textit{Kepler} cross-match\footnote{https://gaia-kepler.fun/} to identify sources with a \textit{Gaia} radial velocity measurement. We generate dynamical age predictions according to the methods described in Section \ref{sec:methods} for 100,885 \textit{Kepler} stars for which we can measure a vertical action. Figure~\ref{fig:KeplerAges} shows combined histograms for \textit{Kepler} stellar age predictions, binned by \teff. For the \textit{Kepler} stellar sample, we have employed the \teffsp values presented in the \textit{Kepler} Data Release 25 (DR25) catalog \citep{mathur_revised_2017}. We obtain the data set from the NASA Exoplanet Archive \citep{kepler_mission_kepler_2019}\footnote{Accessed on 2023-12-05}. We first draw a randomly-chosen 100 points from the stellar age  probability distribution generated by \texttt{zoomies} for each star, before combining these probability samples for all stars within a given \teff\ range. As we expect, we find that the hottest stars are more likely to have younger predicted ages, while age predictions for the coolest stars are more uniformly distributed between young and old. We note that the majority of the \textit{Kepler} stellar samples does not match the mass range of our calibration samples, and so emphasize that these age predictions should be carefully considered as relative age predictions rather than absolute ages. 

\begin{figure}[htbp]
    \centering
    \includegraphics[width=\linewidth]{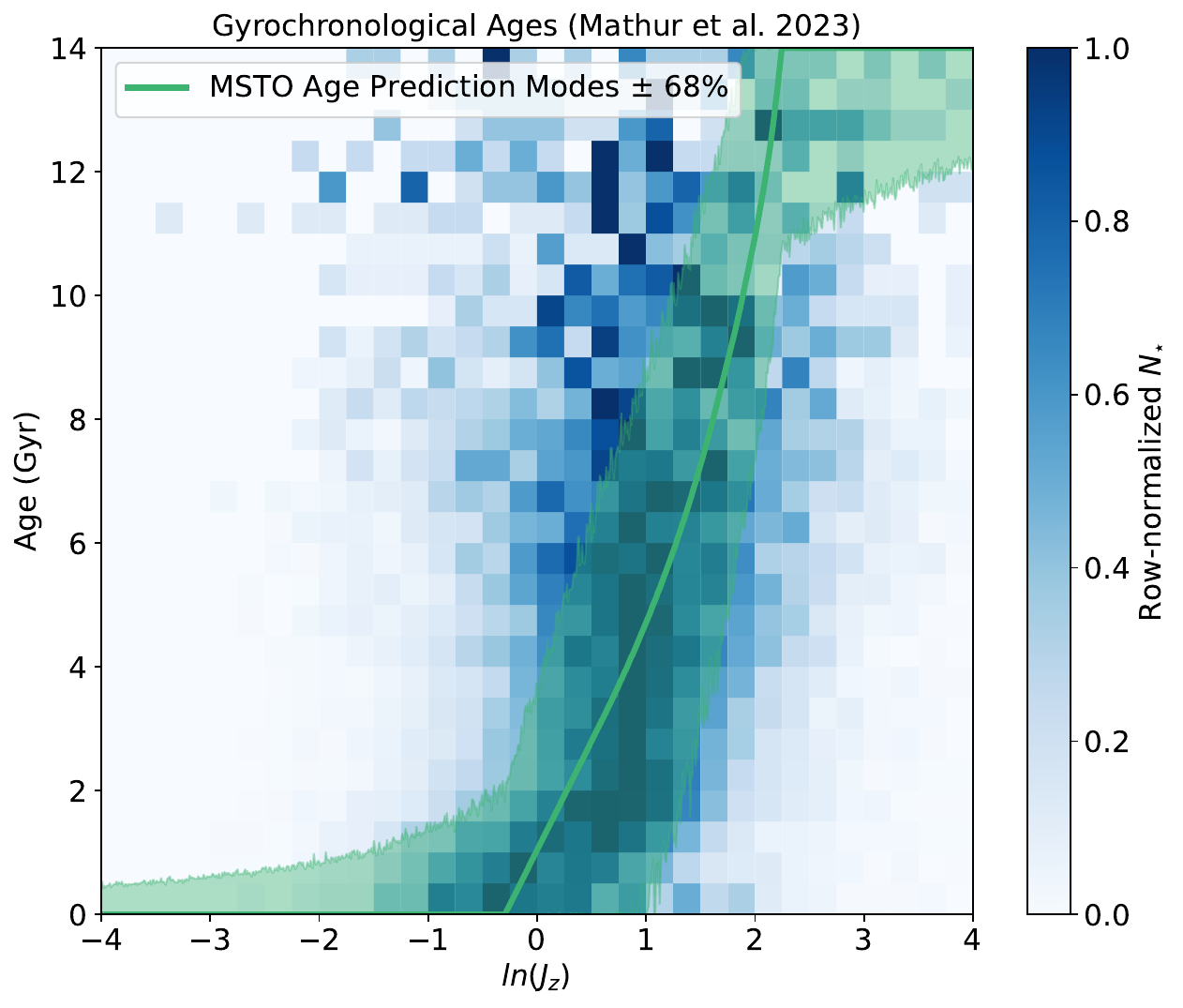}
    \caption{A 2-D histogram of the \citet{mathur_magnetic_2023} sample of \textit{Kepler} main-sequence stars in gyrochronological age-\lnjzsp space. The MSTO-calibrated dynamical ages for this sample are designated with the green line and shaded region. The solid green line represents the statistical mode of each dynamical age probability distribution associated with each star in the \citet{mathur_magnetic_2023} sample, and the shaded region represents $68\%$ of the distribution mass around the mode. The 2-D histogram is row-normalized to reduce the visual effect of the stellar density distribution on the $\tau$ axis.
    }
    \label{fig:Keplergyro}
\end{figure}

We also present a preliminary comparison of our age predictions with the \citet{mathur_magnetic_2023} sample of \textit{Kepler} gyrochronological ages. \citet{mathur_magnetic_2023} measured gyrochronological ages for around 55,000 main-sequence stars using \textit{Kepler}. In Figure \ref{fig:Keplergyro}, we show a comparison of the gyrochronological ages from \citet{mathur_magnetic_2023} with the analogous dynamical age estimates from the MSTO-calibrated model. We consider the MSTO-calibrated age predictions for the Kepler sample because the Kepler sample \lnjzsp distribution matches the \lnjzsp distribution of the MSTO sample more closely than the RGB sample (see Figure \ref{fig:KeplerAges}, right panel). 

While the scatter is large for $\tau > 4$ Gyr in age-action space, the positive relationship between \lnjzsp and age can be seen for young ages in Figure \ref{fig:Keplergyro}. As we expect, the slope of the dynamical age relation is broadly consistent with the slope seen in age--action space for the \citet{mathur_magnetic_2023} \textit{Kepler} gyrochronology sample. With this comparison, we demonstrate that dynamical ages generated with this method are consistent with independently measured gyrochronological ages of main-sequence Kepler stars primarily for  $\tau < 4$ Gyr.

\begin{figure*}
    \centering
    \includegraphics[width=\textwidth]{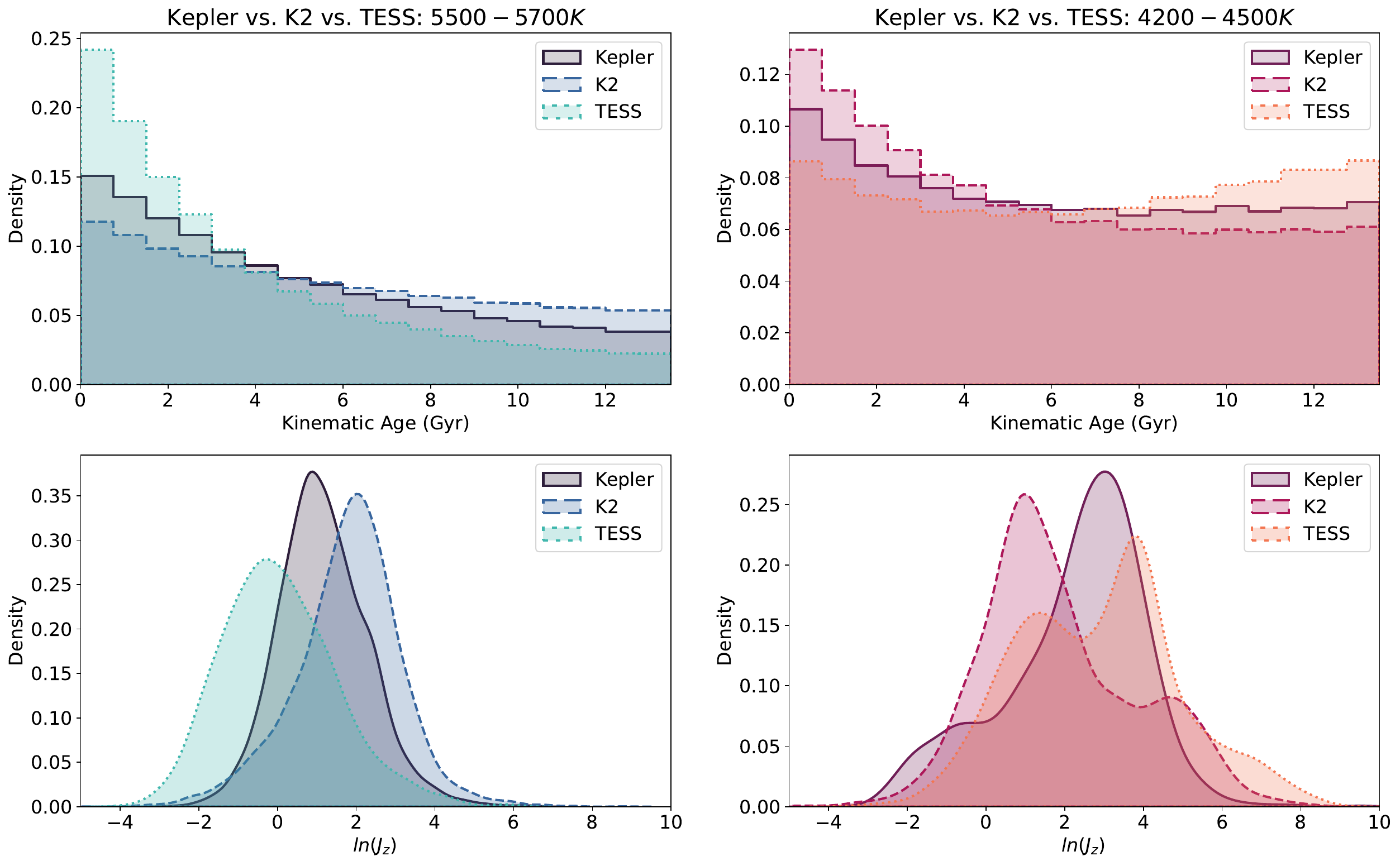}
    \caption{ \textit{Top Left:} Combined dynamical age probability distributions for \textit{Kepler} (solid), K2 (dashed), and \textit{TESS} (dotted) stellar samples with \teffsp between 5500 and 5700 K. Bin widths arbitrarily set to 0.5 Gyr. \textit{Top Right:} Same as top left, but for stars with \teffsp between 4200 and 4500 K. \textit{Bottom Left:} Combined \lnjzsp kernel density estimations (KDEs) for the \textit{Kepler}, K2 and \textit{TESS} stars with \teffsp between 5500 and 5700 K. \textit{Bottom right:} Same as bottom left, but for stars with \teffsp between 4200 and 4500 K.}
    \label{fig:KeplerK2Tess}
\end{figure*}

We compare the ages of the \textit{Kepler}, K2 and \textit{TESS} survey samples. We repeat the process of generating dynamical age estimates for the K2 and \textit{TESS} target stars with positive parallax measurements and radial velocities from \textit{Gaia}. For a consistent comparison, we compare stellar ages within the same selected ranges of \teff. For the K2 stellar sample, we employ the \teffsp values presented in the K2 Ecliptic Plane Input Catalog \citep{huber_k2_2016}, and for the \textit{TESS} stellar sample, we employ the \teffsp values presented in the \textit{TESS} Input Catalog 8 (TIC 8) \citep{stassun_revised_2019}. We present the combined age distributions for each survey sample, along with the vertical action distributions, in Figure \ref{fig:KeplerK2Tess} for two selected temperature ranges. K2 stars appear slightly older on average than \textit{Kepler} stars. Differences in the age distributions between the transit survey samples originate from different mission designs: the surveys' different selection functions, combined with different pointing strategies, contribute to differences in the average observed stellar populations' kinematics. While the original \textit{Kepler} mission's field of view was a small stationary area pointing near, but not directly on, the Galactic plane \citep{gilliland_kepler_2011}, the K2 mission observed several target fields along the ecliptic across a period of several years \citep{howell_k2_2014}. Yet, both missions use the same instrument and bandpass. This may cause different distributions in vertical action, as seen in Figure \ref{fig:KeplerK2Tess}. In contrast, \textit{TESS}'s wide, near-infrared bandpass and its pointing throughout the celestial sphere \citep{ricker_transiting_2014} may be reflected in the difference in age and vertical action distributions between hotter and cooler \textit{TESS} stars, compared to the \textit{Kepler} and K2 samples. We have confined ourselves in this manuscript to a description of the \texttt{zoomies} machinery, and leave its detailed application to exoplanetary studies to future work. However, we comment upon large-scale trends to emphasize the potential usefulness of dynamical ages to the exoplanet community.

\section{Discussion} \label{sec:discussion}

In the following subsections, we return to our assumptions about the dynamical age model's dependence on stellar mass and discuss the limitations of our model related to sample distributions in action space. We also discuss the implications of these results on planet occurrence.

\subsection{Stellar Mass Dependence} \label{subsec:stellarmassdependence}
As discussed in Section \ref{sec:intro} and Section \ref{sec:methods}, we assume in our dynamical age relation that the kinematic galactic motion of stars are mass-independent, consistent with an ``upside-down" and ``inside-out" model of galactic assembly \citep{bird_inside_2013, bird_inside_2021}. Alternatively, it is possible that the AVR is driven primarily by orbital heating (e.g. \citealt{spitzer_possible_1953}). If stellar kinematic properties are determined primarily by orbital heating, the dynamical age model we construct will not be entirely mass-independent and may produce less accurate age predictions for stars with masses significantly different from our calibration sample (i.e. the lowest-mass stars). However, we contend that in this case, this dynamical age relation is still far less dependent on stellar mass than other age determination methods, such as isochrone fitting and asteroseismic ages, especially when interpreting age predictions as relative ages for low-mass stars. In order to avoid any mass dependence in a dynamical age model, a large calibration sample of low-mass stars with relatively confident age measurements is needed.

\subsection{Impact of the full orbital action distribution on age predictions}\label{subsec:allactions}

Here we discuss the extent to which one may apply the dynamical age model to prediction samples that do not closely match the calibration sample in action space. In selecting a calibration sample, we necessarily assume that the \lnjzsp distribution of the calibration sample matches that of the prediction sample of interest. However, we selected the RGB and MSTO samples primarily because of the accuracy and precision of their ages determined with asteroseismology and isochrone-fitting, respectively (not necessarily because their \lnjzsp distributions match the \lnjzsp distributions of our prediction samples). In many cases, sufficiently large calibration samples with accurate, independently-measured ages that also match the \lnjzsp distribution of the prediction sample are not available. This is particularly likely for the lowest-mass stars, which also happen to be the most prominent exoplanet hosts.

We calibrate the dynamical age model in this work on \Jzsp, which represents one slice of a three dimensional projection of action space, including \Jzsp, $J_{r}$, and $J_{\phi}$. We contend that the best way to mitigate the effects of prediction sample that do not match the calibration sample is to build a joint age--action model that considers all three actions. We will address this in a future work, as an age prediction method based on all orbital actions should produce more precise ages. However, short of building a three-dimensional age--action model, one workaround with the current one-dimensional method is to re-sample or re-weight the prediction sample in action space to more closely match the calibration sample.

\subsection{Planet Occurrence and Dynamical Ages}

The investigation of planet formation and evolution is only one application of demographic stellar age constraints. While we aim for \verb|zoomies| to be broadly useful the astronomical community interested in stellar ages in the Milky Way, we focus here on its potential use for studying exoplanet host stars. \textit{Gaia} has enabled exoplanet study to occur within a new Galactic context (see e.g. \citealt{winter_stellar_2020, dai_planet_2021, chen_planets_2021-1, longmore_impact_2021, bashi_quantifying_2021, zink_scaling_2023}). In this Section, we describe the relative strength of dynamically-derived stellar ages for investigations of the (on average) $\mathcal{O}10^{5}-10^{6}$ stars monitored by space-based transit surveys.

Folding together both isochrone or spectroscopic age constraints together with \textit{Gaia} parallaxes or kinematic measurements, the evolution of planetary systems over time has already been seen \citep{berger_gaia-kepler-tess-host_2023, chen_planets_2021}. For example, the radius distribution of planets shifts as stars age \citep{berger_gaia-kepler_2020-1}, and there exists an apparent change in transit multiplicity as well \citep{chen_planets_2021}. \cite{zink_scaling_2023} found that raw occurrence is affected by Galactic height (as determined by \textit{Gaia} parallaxes), one of many proxies for stellar age. We consider the dynamical ages for the population of stars observed by \textit{Kepler}. Nominally, we might expect to see some trend with age-- though dynamical age dating involves much larger uncertainty than other methods. 

\begin{figure}
    \centering
    \includegraphics[width=\linewidth]{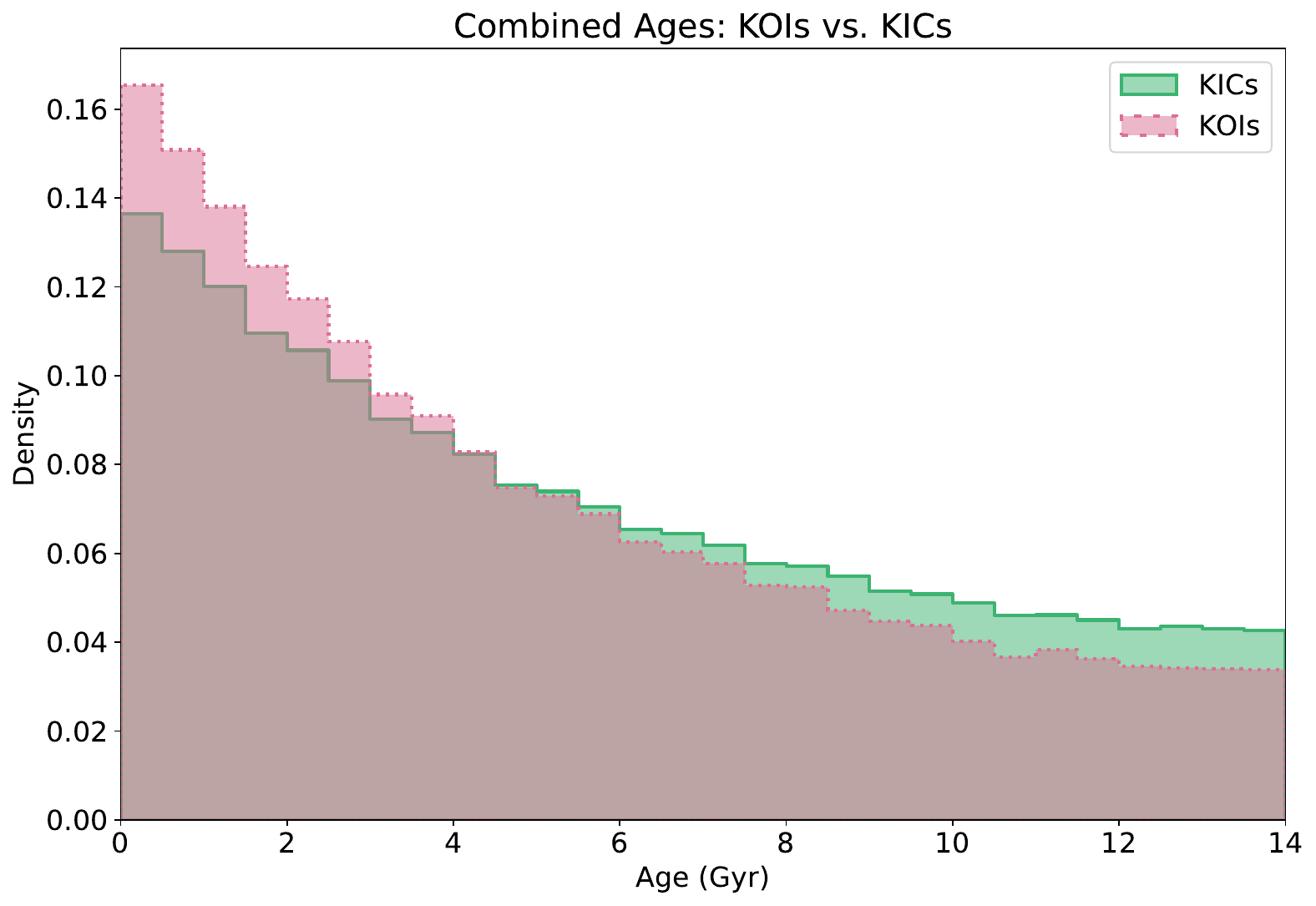}
    \caption{Combined dynamical age distributions for the entire KIC catalog (green) and all KOIs (red).}
    \label{fig:Keplerkoi}
\end{figure}

The degree to which the dynamical age relation we developed is constraining at the population level for the \textit{Kepler} sample is shown in Figure \ref{fig:Keplerkoi}. We overplot the dynamical age distributions for the entire sample of \textit{Kepler} Objects of Interest (KOI), versus KIC field stars with no transiting planets. We do see a visible offset in the same direction as \cite{chen_planets_2021}: that is, host stars to transiting planets are, on average, younger than field stars. This figure is meant to convey the relative strength of dynamical age information alone: fully disambiguating the ages of field stars vs. planet hosts would require many additional steps (given that the population of field stars in fact hosts planets too, though not in a transiting geometry). In this sense, the KIC sample is contaminated with planet hosts and not representative of stars that don't have planets. 
Even given this degree of contamination, Figure \ref{fig:Keplerkoi} shows to first order the extent to which dynamical information alone can be useful at a demographic level. 

\section{Conclusions} \label{sec:conclusions}

We summarize our results and work as follows:
\begin{itemize}
\item We calibrate a flexible model to predict stellar ages given the vertical actions of stars using two independent calibration samples: one based on asteroseismic red giant branch star ages and one using isochrone main-sequence turn-off star ages.
\item We compare these dynamical age predictions with an external sample of red giant branch stars with asteroseismic and spectroscopic ages and find general agreement, though our age predictions are less precise (as expected).
\item We compare our dynamical age predictions with measured ages for open clusters and main-sequence stars with gyrochronological ages and again find that our ages agree with other independent methods, at least in relative age ranking.
\item We produce age predictions for the \textit{Kepler}, K2, and \textit{TESS} main samples and show that the stars observed by these surveys have naturally different Galactic kinematics and intrinsic age distributions.
\item We release our model and the age prediction tools used in this analysis in an open-source, user-friendly software package called \texttt{zoomies} \citep{sagear_ssagearzoomies_2024}.
\end{itemize}

The magnitude and precision of \textit{Gaia} data that have become available in recent years allows us to generate homogeneous, mass-independent relative age predictions using only astrometric data and a potential model of the Milky Way. This newly enables large-scale age analyses on a wide range of stellar masses, even the lowest-mass stars. These dynamical ages can be useful for any investigation involving stellar populations, and especially for demographic studies of the lowest-mass exoplanet hosts and the dynamical evolution of their planetary systems. In a future work, we plan to combine orbital eccentricity information for planet-hosting stars \citep[e.g.][]{van_eylen_orbital_2019, sagear_orbital_2023} with homogeneous stellar metallicity information \citep[e.g.][]{anderson_higher_2021} to disentangle the underlying relationship between planet occurrence, orbital dynamics, metallicity, and age.

We thank the anonymous referees for helpful suggestions that improved the quality of this work. We thank Jamie Tayar and Zachary Claytor for careful readings of this manuscript, along with Christopher Lam, Quadry Chance, Natalia Guerrero, Dana Yaptangco, Nazar Budaiev, and Soichiro Hattori for helpful discussions about this analysis.

This work has made use of data from the European Space Agency (ESA) mission {\it Gaia} (\url{https://www.cosmos.esa.int/gaia}), processed by the {\it Gaia} Data Processing and Analysis Consortium (DPAC, \url{https://www.cosmos.esa.int/web/gaia/dpac/consortium}). Funding for the DPAC has been provided by national institutions, in particular the institutions participating in the {\it Gaia} Multilateral Agreement. 
This paper includes data collected by the Kepler, K2, and TESS missions and obtained from the MAST data archive at the Space Telescope Science Institute (STScI). The specific observations analyzed can be accessed via \dataset[https://doi.org/10.17909/T9059R]{https://doi.org/10.17909/T9059R} (KIC), \dataset[https://doi.org/10.17909/T93W28]{https://doi.org/10.17909/T93W28} (EPIC), and \dataset[https://doi.org/10.17909/fwdt-2x66]{https://doi.org/10.17909/fwdt-2x66} (TIC). STScI is operated by the Association of Universities for Research in Astronomy, Inc., under NASA contract NAS5–26555. Support to MAST for these data is provided by the NASA Office of Space Science via grant NAG5–7584 and by other grants and contracts. This research has made use of the NASA Exoplanet Archive, which is operated by the California Institute of Technology, under contract with the National Aeronautics and Space Administration under the Exoplanet Exploration Program.

\vspace{5mm}
\facilities{Gaia, Kepler, TESS, Exoplanet Archive, MAST}

\software{numpy \citep{harris_array_2020}, matplotlib \citep{caswell_matplotlibmatplotlib_2024}, astropy \citep{robitaille_astropy_2013, collaboration_astropy_2018, collaboration_astropy_2022}, numpyro \citep{bingham_pyro_2019,phan_composable_2019}, gala \citep{price-whelan_gala_2017, price-whelan_adrngala_2020}, agama \citep{vasiliev_agama_2019}}

\bibliography{main}{}
\bibliographystyle{aasjournal}

\appendix

\section{Relative Age Comparisons}\label{appendix:relative}

\begin{figure}[htbp]
    \centering
    \includegraphics[width=\linewidth]{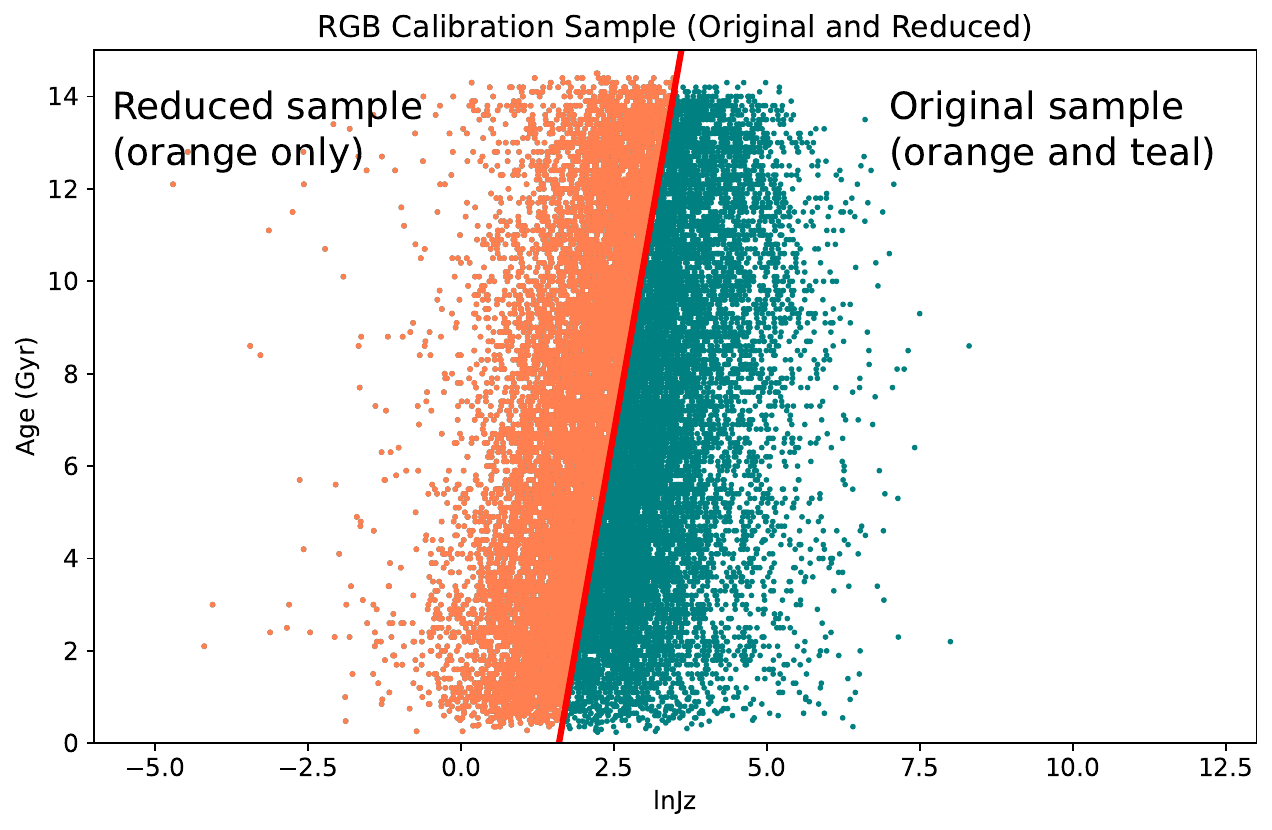}
    \caption{Original and reduced RGB calibration samples. The original RGB calibration sample is represented by all points (both orange and teal). The orange points only represent the reduced calibration sample: stars in the RGB calibration sample that lie left of the line $age = 7.5 \times \lnjz -12$ (red).}
    \label{fig:RangeTestSample}
\end{figure}

Throughout this work, we state that dynamical stellar age predictions may be interpreted as absolute ages as long as the calibration sample \lnjzsp distribution overlaps with the prediction sample \lnjzsp distribution. We perform a preliminary check for this statement.  We begin with artificially reducing the RGB calibration sample by removing all stars to the right of the line $age = 7.5 \times \lnjz -12$. We do this to ensure that the natural slope of the age–\lnjzsp relation is preserved, while ensuring only the smallest \lnjzsp values remain in this calibration sample. The original and reduced calibration samples are shown in Figure \ref{fig:RangeTestSample}.

\begin{figure}[htbp]
    \centering
    \includegraphics[width=\linewidth]{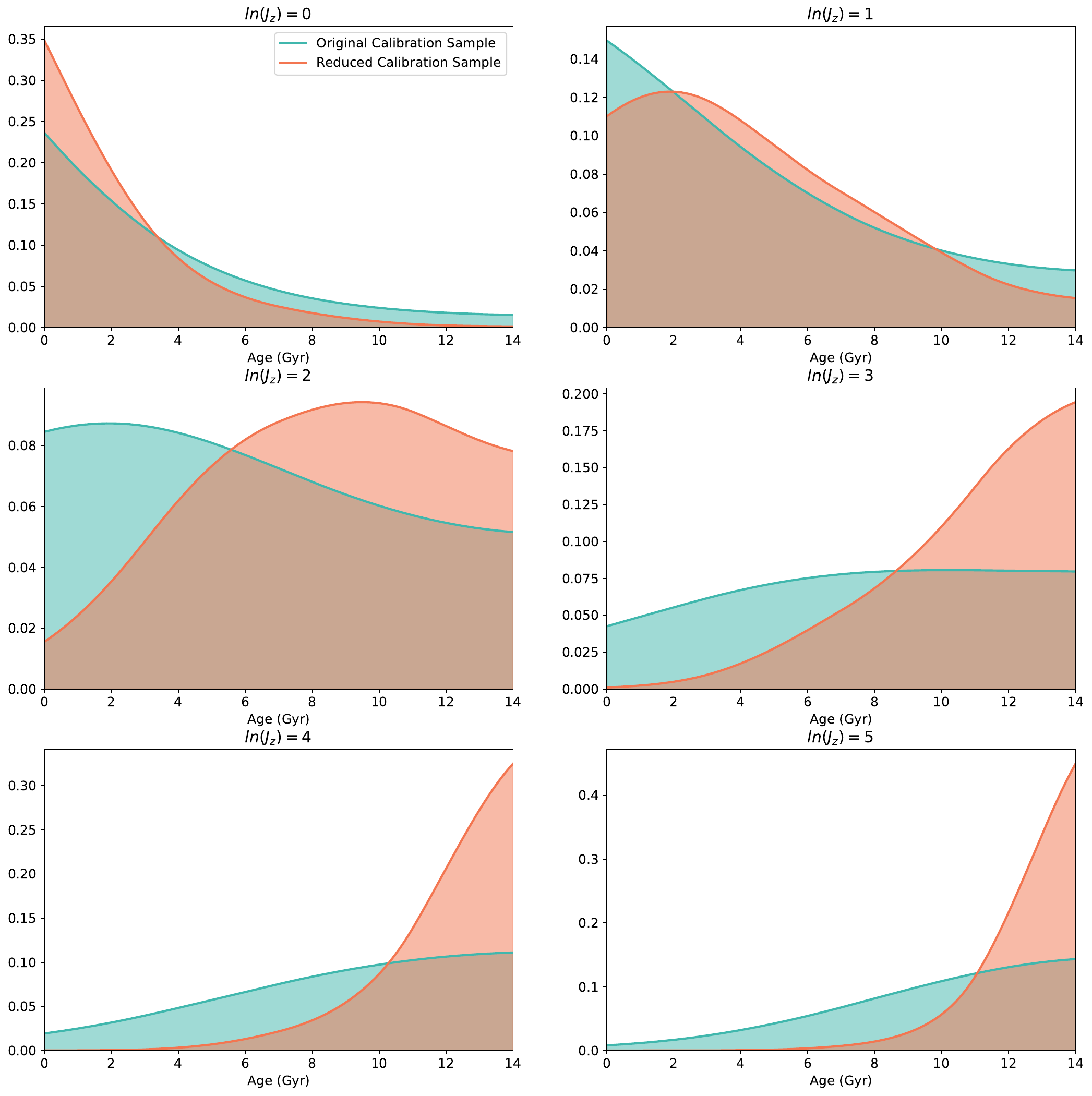}
    \caption{Comparison between original RGB age predictions (orange) and reduced RGB age predictions (teal) for six test values of \lnjz. For the values of \lnjzsp which are well-represented by both the original and reduced calibration sample ($\lnjz = 0$ and $\lnjz = 1$), both age predictions appear similar. For larger values of \lnjzsp which are not well-represented in the reduced calibration sample, the absolute age predictions differ more significantly, while the relative age rankings remain consistent.}
    \label{fig:RangeTestPrediction}
\end{figure}

We then perform the age--\lnjzsp spline calibration process described in Section \ref{sec:methods} using the reduced calibration sample. We generate age predictions for six selected \lnjzsp values using the new reduced age–\lnjzsp spline model, and we compare these age predictions with age predictions generated with the original age–\lnjzsp spline model. In Figure \ref{fig:RangeTestPrediction}, we show a comparison of the two age predictions for six values of \lnjzsp.

We do find that the age predictions appear similar for \lnjzsp values that are represented in both calibration samples (\lnjzsp values of 0 and 1), but they differ for \lnjzsp values where only one calibration sample is defined (\lnjzsp values of 2 and above). However, as expected, the relative ages match (i.e. a larger \lnjzsp value results in an overall older age prediction in both cases).

\section{Impact of Galactic Models on Age Prediction}\label{appendix:galmodels}

It is necessary to choose a galactic potential model in order to calibrate the age--action relation, as described in Section \ref{sec:methods}. We perform a preliminary test to check the impact the choice of galactic model has on stellar age predictions. We compare age predictions for three extreme choices for galactic model, where the mass of the galaxy is composed entirely of a disk, bulge, or halo. We compare these resulting age predictions to the age predictions generated by the Milky Way potential model and the RGB calibration sample used in this work. We calibrate three age–action spline models using the method described in Section \ref{sec:methods}, but changing only the galactic models. For the ``All Disk” model, the galactic model is composed of a disk of $10^{10} M_{\odot}$, and no halo or bulge. For the ``All Bulge” model, the galactic model is composed of a bulge of $50^{9} M_{\odot}$, and no disk or halo. For the ``All Halo” model, the galactic model is composed of a halo of $10^{11} M_{\odot}$, and no disk or bulge.

\begin{figure}[htbp]
    \centering
    \includegraphics[width=\linewidth]{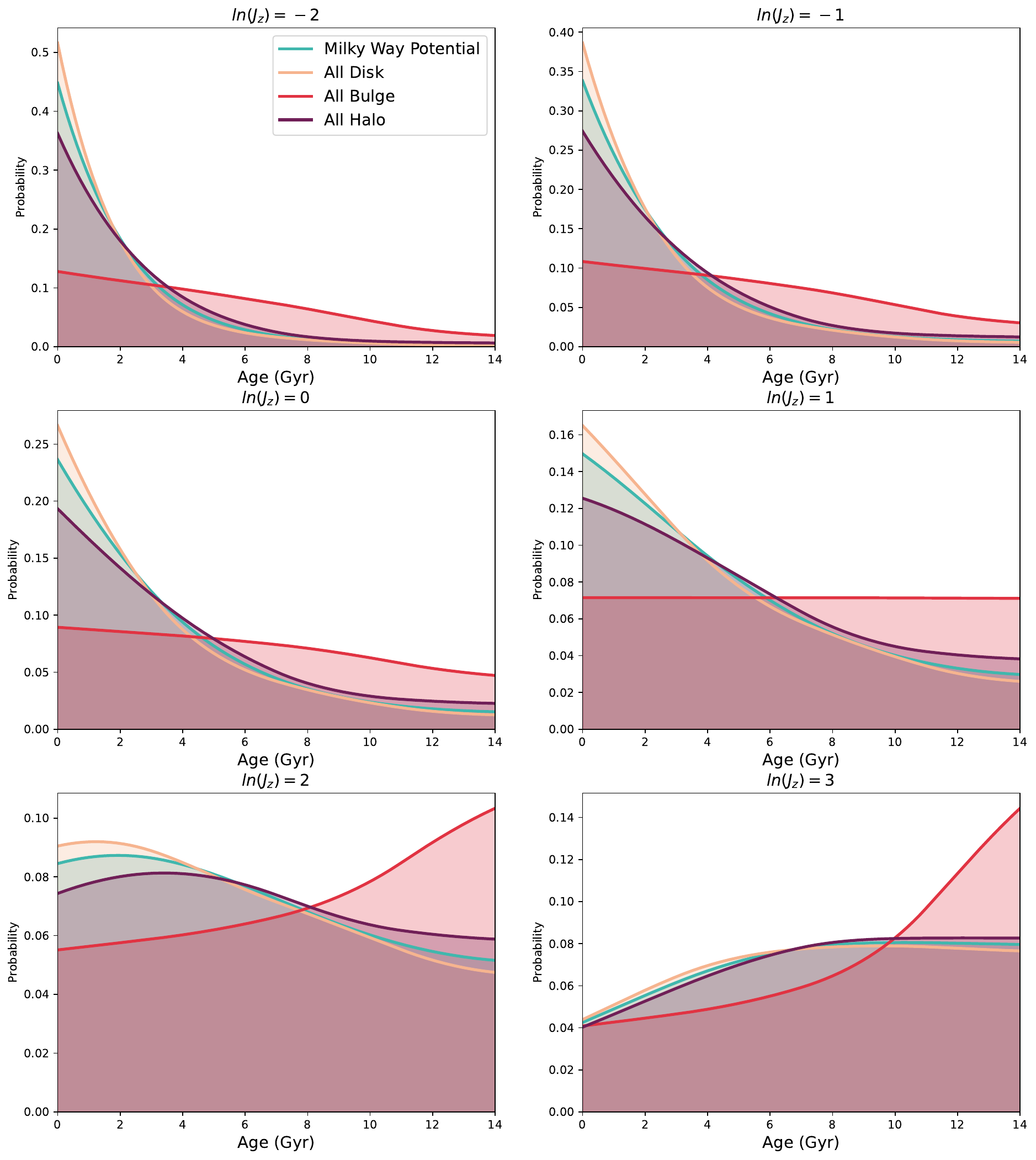}
    \caption{Comparison between age predictions (using the RGB calibration sample) generated with the Milky Way potential model (blue) vs. extreme custom galactic models, where the galaxy is composed entirely of a massive disk (orange), bulge (red), or halo (purple). Age predictions are shown for six test values of \lnjzsp between $-2$ and $3$.}
    \label{fig:galmodels}
\end{figure}

We choose six \lnjzsp test values between $-2$ and $3$ to generate and compare age probability distributions for each model. In Figure \ref{fig:galmodels}, we show the age predictions using all three ''test” galactic models, and the original Milky Way model used in the paper including a disk, bulge, and halo. We find that the entirely-disk and entirely-halo galactic potential models (orange and purple) produce age probability distributions very similar to that produced by the Milky Way model. The entirely-bulge potential model (red) differs more significantly, but critically, the ages rank appropriately (i.e. the age probability distributions always increase with increasing values of \lnjz, no matter which galactic model is used).

We choose the most extreme galactic models for this demonstration, but in practice, it would be reasonable to choose a galactic potential model that resembles the Milky Way model. Therefore, we conclude that any reasonable choice of galactic model will at most slightly affect the age probability distributions. Even more extreme choices of galactic models are not likely to significantly impact stellar age predictions.

\section{Dependence of Intrinsic Age Scatter on Stellar Parameters}\label{appendix:intrinsicscattercheck}

\begin{figure}[htbp]
    \centering
    \includegraphics[width=0.7\linewidth]{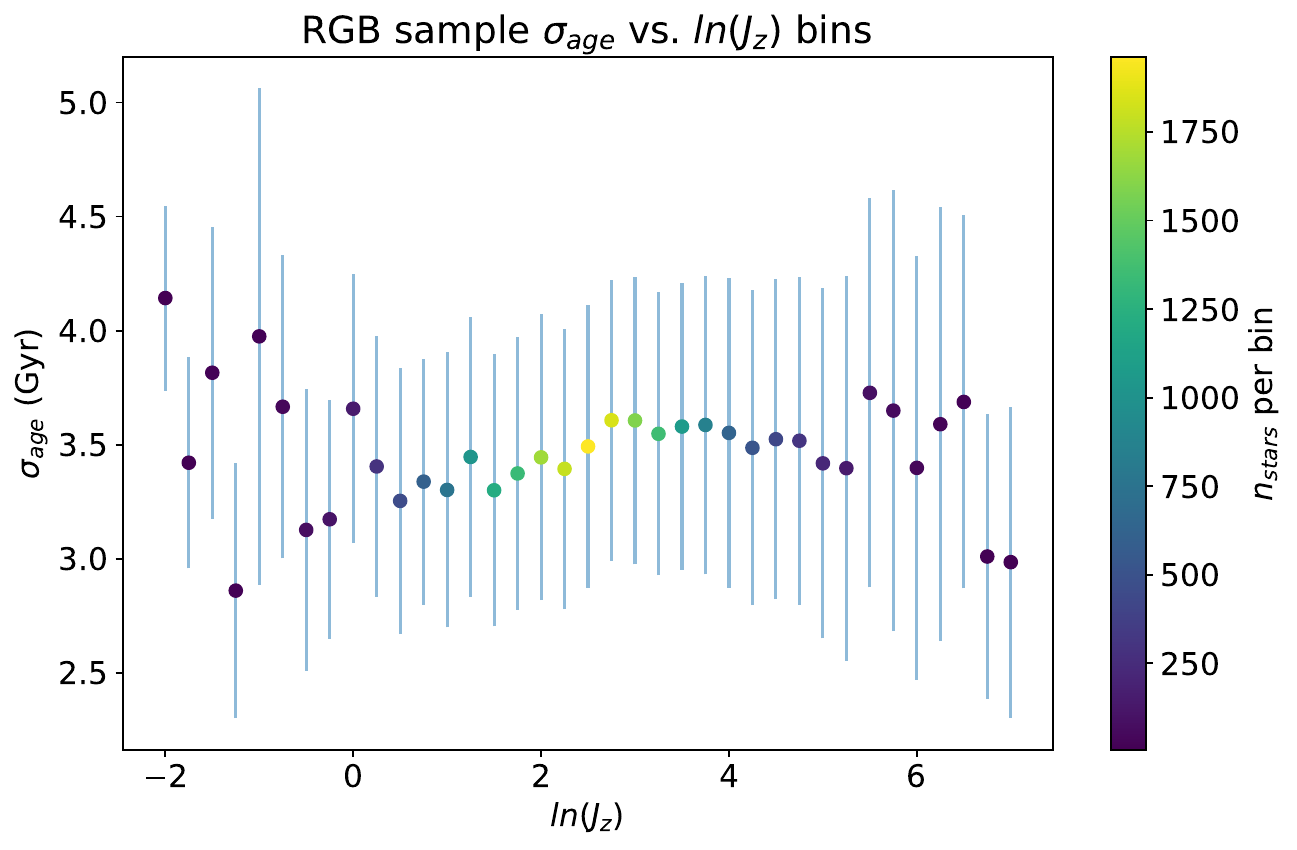}
    \caption{Standard deviation of stellar ages in 37 bins of \lnjzsp vs. midpoint of each \lnjzsp bin for the RGB calibration sample. The error bars on $\sigma_{\tau}$ represent the standard deviation of individual stellar age uncertainties for the stars in each bin. The color represents the number of stars included in each bin; the small sample size for bins with \lnjzsp $<0$ and $>5$ likely contributes to the large scatter in $\sigma_{\tau}$ for low and high \lnjzsp bins.}
    \label{fig:intrinsicscatterappendix}
\end{figure}

We conducted a preliminary test to check for any clear correlations between the intrinsic calibration age scatter and several stellar age parameters, such as stellar metallicity, $log(g)$, alpha abundance, mass, radius, distance, and effective temperature. We begin by splitting up the RGB calibration sample into 37 arbitrary bins of \lnjz, and calculating the standard deviation of stellar (calibration) ages within each bin. This metric serves as an approximation for the intrinsic scatter in age within each \lnjzsp bin. In Figure \ref{fig:intrinsicscatterappendix}, we show the standard deviation of stellar ages, $\sigma_{\tau}$, for the stars in each \lnjzsp bin vs. the midpoint of each \lnjzsp  bin.

\begin{figure}[htbp]
    \centering
    \includegraphics[width=0.8\linewidth]{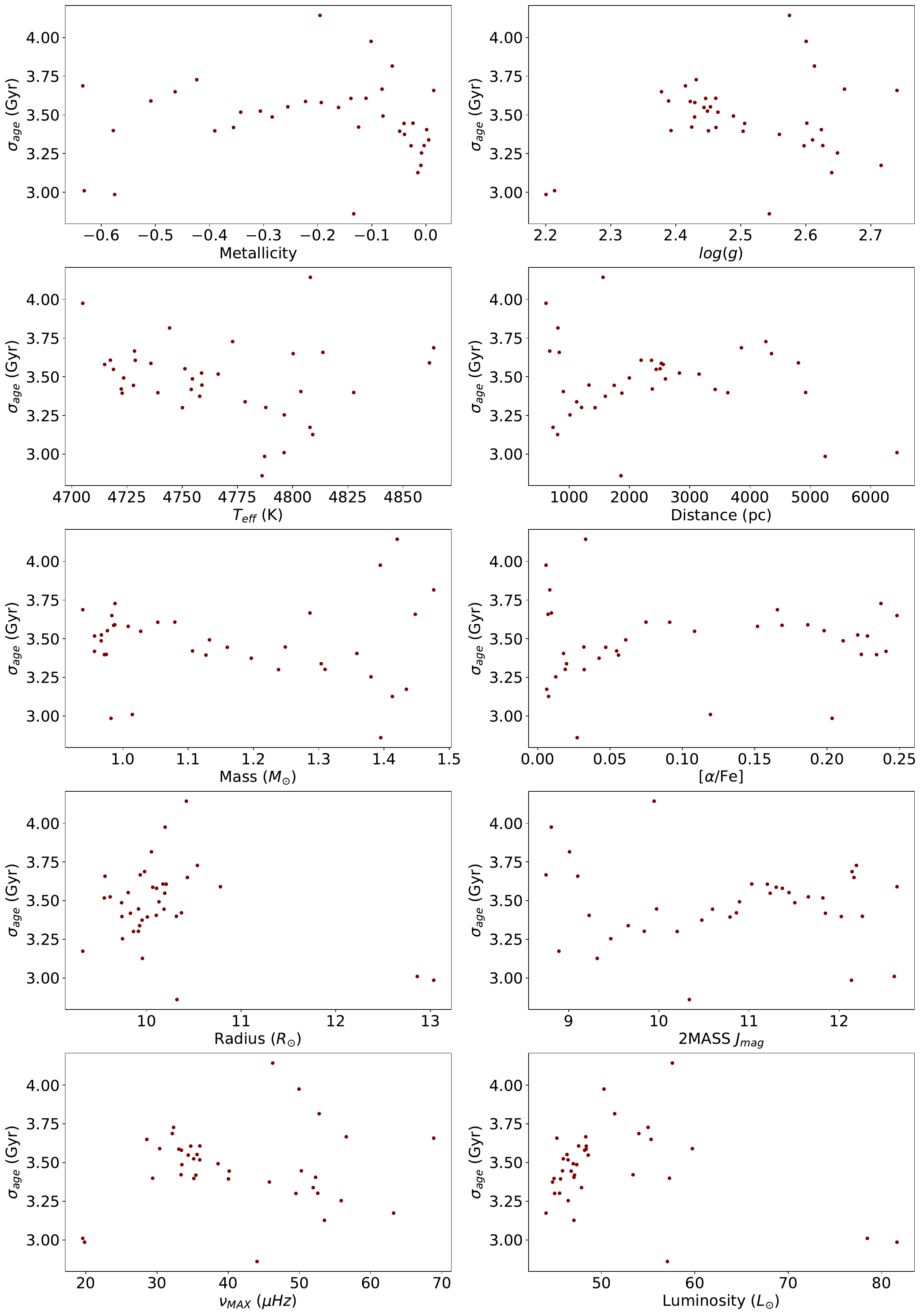}
    \caption{Standard deviation of RGB calibration ages in 37 \lnjzsp bins vs. median stellar parameter values in each bin (stellar metallicity, $log(g)$, $T_{\mathrm{eff}}$, distance, stellar mass, stellar radius, $\alpha$/Fe, $J$-band magnitude, maximum frequency $\nu_{\mathrm{max}}$, and luminosity). All stellar parameters are published alongside the RGB calibration ages in \citet{stokholm_unified_2023}.}
    \label{fig:intrinsicscatterstellarparams}
\end{figure}

We then calculate the median of several stellar parameters for the calibration stars in each \lnjzsp bin. We use the stellar parameters published alongside the calibration ages in \citet{stokholm_unified_2023}. We plot $\sigma_{\tau}$ against the median stellar parameter value for ten different stellar parameters, shown in Figure \ref{fig:intrinsicscatterstellarparams}. We do not find evidence for any clear correlations between intrinsic age variance and these stellar parameters. Indeed, action-driven ages are uniquely useful because they are based on stellar kinematics, and the dynamical age uncertainty is dominated by the intrinsic variance in the age calibration sample. Thus, the uncertainty in dynamical age predictions have little dependence on individual stellar properties.

\section{Open Cluster Age Comparisons}\label{appendix:cluster}

\begin{figure}[htbp]
    \centering
    \includegraphics[width=\linewidth]{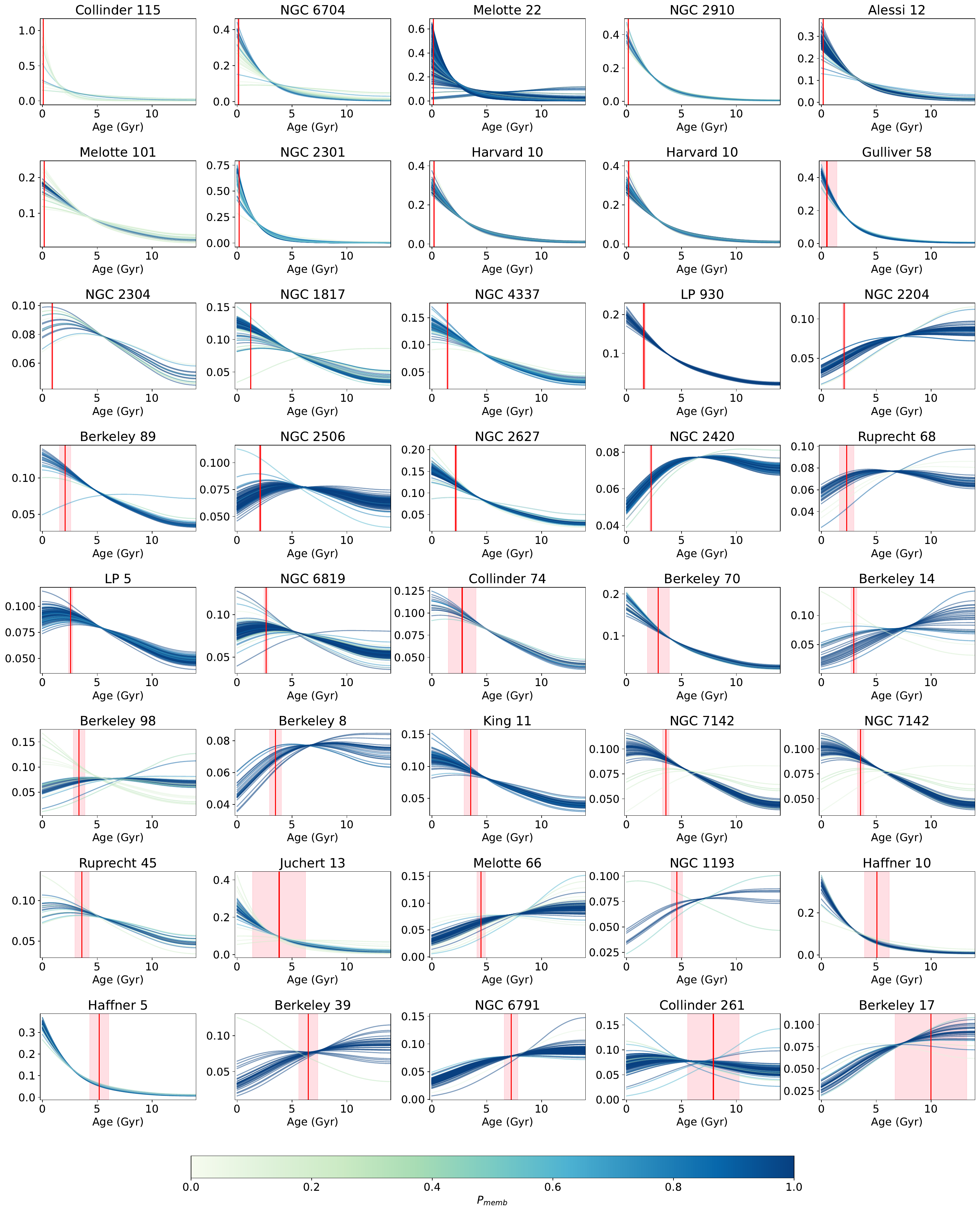}
    \caption{Age probability distributions for randomly selected open clusters from the \citet{dias_updated_2021} sample. 15 clusters with $0 < \tau < 2$ Gyr and 25 clusters with $2 < \tau < 14$ Gyr according to \citet{dias_updated_2021} were randomly selected to present a mix of young and old clusters. Each subplot shows the age probability distributions for each star associated with each cluster, with color denoting the membership probability according to \citet{dias_updated_2021}. The red vertical lines and shaded regions mark the cluster ages and $1\sigma$ uncertainties (respectively) presented in \citet{dias_updated_2021}.}
    \label{fig:clustergrid}
\end{figure}

We present an extension of the open cluster dynamical age comparison described in Section \ref{subsec:openclusterages}. We consider the open clusters listed in \citet{dias_updated_2021} with more than 100 listed members. Since the vast majority of open clusters in \citet{dias_updated_2021} have $\tau < 1$ Gyr, we randomly select 15 clusters with $0 < \tau < 2$ Gyr and 25 clusters with $2 < \tau < 14$ Gyr according to \citet{dias_updated_2021}, with uniform probability within each bin. For each star associated with each chosen open cluster, we calculate \lnjzsp and the corresponding dynamical age probability distribution using the RGB-calibrated dynamical age model. In each subplot of Figure \ref{fig:clustergrid}, we show the probability distribution of each star associated with each cluster (teal), along with the cluster age and $1\sigma$ age uncertainty presented in \citet{dias_updated_2021} with the red lines and shaded regions, respectively.

\end{document}